\documentclass{article}

\usepackage{graphics} 
\usepackage{epsfig} 
\usepackage{mathptmx} 
\usepackage{times} 
\usepackage{amsmath} 
\usepackage{amssymb}  
\usepackage{subcaption}
\usepackage{bm}
\usepackage{color}
\usepackage{newtxtext, newtxmath}
\usepackage{dblfloatfix}

\newcommand{\vecA}{\mathbf{A}}
\newcommand{\vecB}{\mathbf{B}}

\newcommand{\vecF}{\mathbf{F}}

\newcommand{\vecH}{\mathbf{H}}
\newcommand{\vecI}{\mathbf{I}}
\newcommand{\vecJ}{\mathbf{J}}

\newcommand{\vecM}{\mathbf{M}}

\newcommand{\vecQ}{\mathbf{Q}}
\newcommand{\vecR}{\mathbf{R}}

\newcommand{\veca}{\mathbf{a}}

\newcommand{\vecq}{\mathbf{q}}
\newcommand{\vecs}{\mathbf{s}}
\newcommand{\vecv}{\mathbf{v}}

\title{\LARGE \bf
Planar Multi-link Swimmers: Experiments and Theoretical Investigation Using ``Perfect Fluid'' Model*
}

\author{Evgenia Virozub, Oren Wiezel, Alon Wolf and Yizhar Or$^{1}$
\thanks{*This work has been supported by the Israeli
Science Foundation under Grant 567/14 and Technion Autonomous Systems Program grant no. 2021776.}
\thanks{$^{1}$E. Virozub, O. Wiezel, A. Wolf and Y. Or are with the Faculty of Mechanical Engineering, Technion - Israel Institute of Technology, Haifa 32000, Israel.
        Corresponding author Y. Or, {\tt\small izi@technion.ac.il}}%
}

\begin{document}

\maketitle
\thispagestyle{empty}
\pagestyle{empty}
\begin{abstract}
Robotic swimmers are currently a subject of extensive research and development for several underwater applications. Clever design and planning must rely on simple theoretical models that account for the swimmer's hydrodynamics in order to optimize its structure and control inputs. In this work, we study a planar snake-like multi-link swimmer by using the ``perfect fluid'' model that accounts for inertial hydrodynamic forces while neglecting viscous drag effects. The swimmer's dynamic equations of motion are formulated and reduced into a first-order system due to symmetries and conservation of generalized momentum variables. Focusing on oscillatory inputs of joint angles, we study optimal gaits for 3-link and 5-link swimmers via numerical integration. For the 3-link swimmer, we also provide a small-amplitude asymptotic solution which enables obtaining closed-form approximations for optimal gaits. The theoretical results are then corroborated by experiments and motion measurement of untethered robotic prototypes with 3 and 5 links floating in a water pool, showing a reasonable agreement between experiments and the theoretical model.
\end{abstract}

\section{Introduction}

Autonomous swimming robots have a promising potential for various applications such as surveillance and protection in marine environment, search and rescue missions, and maintenance operations within pipe systems of complex infrastructures \cite{vaidyanathan2000hydrostatic,kwak2016design,li2016bio,bayat2016envirobot}. A leading biologically-inspired concept of articulated mobile robots is a snake-like kinematic chain that undergoes body undulations of a travelling wave where the joint angles undergo phase-shifted oscillatory motion \cite{kelasidi2016innovation,mcisaac2002experimental,crespi2013salamandra,ijspeert2014biorobotics}. Coordination between the links and optimization of the gait of periodic shape changes is highly crucial for generating effective net motion.
Terrestrial snakes whose motion is governed by rigid-body contact mechanics have been widely explored for several decades \cite{hirose1993biologically,gong2016kinematic,onal2013autonomous}. On the other hand, the motion of swimming snake robots is governed by hydrodynamic interaction between the fluid and the robots. Several theoretical models of the hydrodynamics of swimming have been studied, with varying level of accuracy and computational complexity. Some works use coefficients of lift and/or drag forces, which can be tuned empirically \cite{mcisaac2002experimental,morgansen2002trajectory,crespi2008online,porez2014improved,kelasidi2014modeling,sfakiotakis2007biomimetic}.
Other works consider the interaction of the swimmer with vortices shed by the undulating tail \cite{tallapragada2015self,zhu2002three}.
Some of the works above consider open-loop periodic inputs of joint angles \cite{mcisaac2002experimental,crespi2008online,porez2014improved,kelasidi2014modeling}, whereas others focus on feedback control and mechanical actuation of joint torques \cite{morgansen2002trajectory,sfakiotakis2007biomimetic}. 
Nonetheless, all the modelling methods mentioned above result in a complicated nonlinear system of second-order differential equations which have to be integrated numerically. The solutions depend strongly on the empirically-tuned drag coefficients. In addition, the resulting motion is time-dependent which does not necessarily reach a steady-state periodic solution, and depends significantly on the frequency of the oscillating inputs. This makes the analysis complicated, as well as sensitive to many hand-tuned parameters.

A fundamentally different formulation approach which results in a remarkably simpler model, is that of ``perfect fluid'' \cite{kanso2005locomotion,melli2006motion,lee2009dynamics}, which assumes inviscid irrotational potential flow, where the swimmer-fluid interaction is induced by reactive forces that represent \textit{added mass effect}, associated with the momentum required in order to displace the fluid surrounding the swimmer's links \cite{lamb1945hydrodynamics}. Using this model, invariance of the dynamics under rigid-body transformation enables reduction into a system of first-order differential equations, which relate the swimmer's body motion to the velocities of shape variables (i.e. joint angles), which are assumed to be directly prescribed.
Importantly, the reduced system is \textit{time invariant}, that is, the motion's time rate simply scales with frequency of the periodic input. Thus,  the net body motion per period depends only on the gait's trajectory and not on its frequency.
A similar ``principal kinematic'' structure of the dynamic equations also holds for other locomotion systems such as wheeled vehicles \cite{ostrowski1996gait,shammas2007geometric} and micro-swimmers in Stokes flow \cite{gutman2016symmetries,hatton2013geometric}. Such systems are widely studied in the robotics literature, using methods of differential geometry and notions of Lie groups \cite{ostrowski1998geometric,kelly1995geometric}. Most of previous works in this field have studied gait planning for achieving desired net motion, which is computed  by using numerical integration \cite{shammas2007geometric} or by applying approximate area-integral rules \cite{hatton2011geometric,hatton2015nonconservativity}. Optimization of gaits for achieving maximal displacement or energetic efficiency has also been studied, and mainly involved numerical computations \cite{cortes2001optimal,ostrowski2000optimal,tam2007optimal,ramasamy2017geometric}. Finally, while several theoretical models of robotic swimming models have been tested experimentally \cite{mcisaac2002experimental,morgansen2002trajectory,porez2014improved,kelasidi2015experimental}, the low-dimensional ``perfect fluid model'' has not yet been validated experimentally.

The goal of this work is to revisit the ``perfect fluid'' model for planar multi-link swimmers and analyze it both theoretically and experimentally. The ``perfect fluid'' model gives a low-dimensional principal kinematic time-invariant  system which depends on very few physical parameters, in contrast to more complicated previous models \cite{mcisaac2002experimental,morgansen2002trajectory,crespi2008online,porez2014improved}. This enables explicit closed-form analysis of the robot's motion under open-loop inputs of periodic gaits, in contrast to previous works that used numerical integration only \cite{shammas2007geometric,hatton2011geometric,hatton2015nonconservativity,cortes2001optimal}. Focusing on small-amplitude harmonic inputs of joint angles, we use perturbation expansion \cite{nayfeh2008perturbation,wiezel2016optimization} in order to obtain asymptotic expressions for the net motion of the three-link swimmer. These expressions enable analysis and optimization of joint angles' stroke amplitude and relative phase, as well as links' length ratio, for achieving maximal net displacement. Additionally, explicit expression for the curvature of net motion as a function of angles oscillation offset is obtained, which enables simple generation of moderate turning motions. For the five-link swimmer, optimization of stroke amplitude and phase difference between consecutive joints is conducted numerically, and a global optimizer is obtained. Validity of the ``perfect fluid'' model is tested by conducting controlled motion experiments of untethered floating prototypes of the three- and five- link swimmers. The experimental and theoretical results are compared by using motion measurements from an optical tracking system. Good qualitative and reasonable quantitative agreement is obtained, after calibrating the added mass effect to account only for the submerged part of the robot's links. Additionally, experimental results that demonstrate optimal phase difference between joints are also shown. This study thus proves the usefulness of the ``perfect fluid'' model as a simplified theoretical tool for studying the dynamics, control and gait optimization of swimming robots. The paper is organized as follows. The next section presents the problem statement and formulation of the dynamic equations. Section III includes asymptotic analysis of the three-link swimmer. Section IV contains numerical simulations and optimization of gaits for three- and five-link swimmers. Section V presents experimental results, and section VI discusses their comparison with prediction of the theoretical model. The closing section summarizes the results and lists possible directions for future extensions of the research. In order to make our analysis accessible to a broader audience of the robotics research community, we chose not to use advanced notions of geometric mechanics such as Lie groups and Riemannian geometry as in previous works \cite{ostrowski1998geometric,kelly1995geometric,hatton2015nonconservativity}. Instead, the swimmer's dynamics is formulated using elementary terminology of linear algebra, vector calculus, and ordinary differential equations.

\section{Problem formulation}

\begin{figure}[!t]
      \centering
      \begin{subfigure}[t]{0.4\textwidth}
      \includegraphics[width=\textwidth]{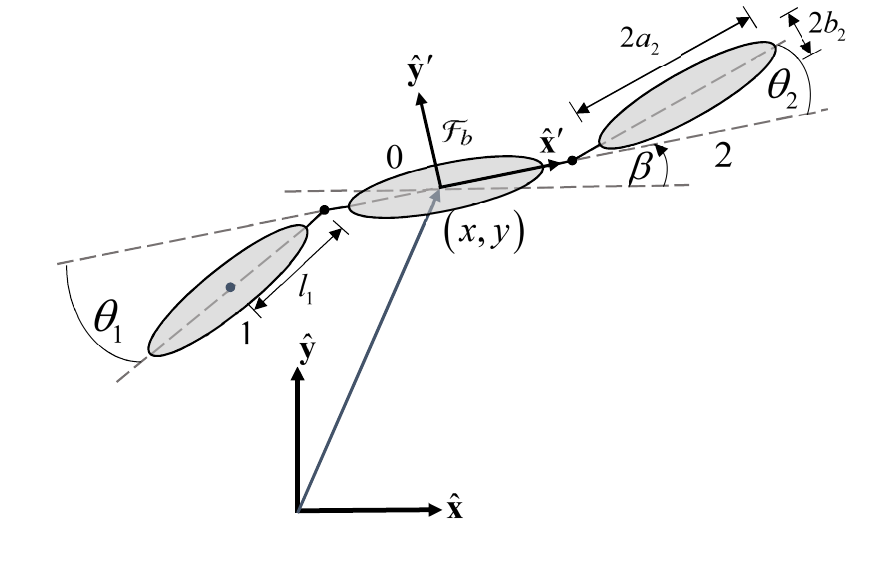}
      \caption{}
      \label{fig:3-link}
      \end{subfigure}
      \hspace{15pt}
      \begin{subfigure}[t]{0.54\textwidth}
      \includegraphics[width=\textwidth]{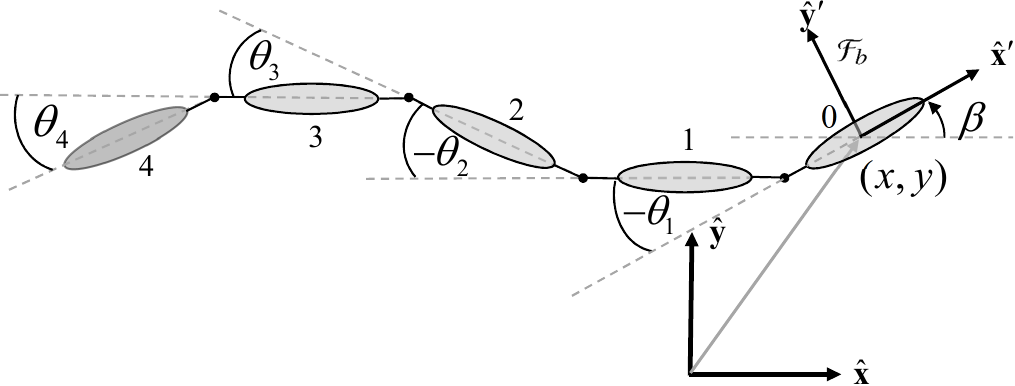}
      \caption{}
      \label{fig:5-link}
      \end{subfigure}
      \caption{Swimmer models. $(x,y)$ are the position of the body-fixed reference frame origin. $\beta$ is the rotation angle of the reference frame. $a_i$ and $b_i$ are the major and minor radii of the elliptic links.  (a) 3-link swimmer model. (b) 5-link swimmer model.}
      \label{fig:models}
   \end{figure}

We now describe the theoretical model of the swimmer and formulate its dynamic equations of motion using the ``perfect fluid'' hydrodynamic model. The planar swimmers shown in Figs. \ref{fig:3-link} and \ref{fig:5-link} respectively,  consist of $N=3$ and $N=5$ links connected by revolute joints. The swimmers' motion is restricted to translation in $(x,y)$ plane and rotations about $z$ axis. Each link is an ellipse with principal radii of $a_i,b_i$ and density $\rho$, that has mass $m_i$ and moment of inertia $I_i$. In order to avoid collisions between adjacent links, the distance between the center of the $i$th link and the adjacent joint is $l_i > a_i$. The relative angles between links are denoted by $\theta_i$. The swimmer is submerged in an unbounded domain of ideal fluid with density $\rho$. That is, the swimmer is neutrally buoyant and gravity effects are not considered. It is assumed that the joint angles are directly controlled, and undergo harmonic oscillations of the form 
\begin{equation}
\theta_i (t)=Asin(\omega t+\varphi_i). 
\label{eq:input}
\end{equation} 
In order to formulate the dynamic equations that govern the swimmer's motion, generalized coordinates are chosen as $\vecq=(\vecq_b,\vecq_s)$, where the body coordinates $\vecq_b=(x,y,\beta)$ describe the position and orientation of a body-fixed frame $\mathcal{F}_b$ attached to link number `$0$', while the shape coordinates $\vecq_s=(\theta_1,\ldots,\theta_{N-1})$ are the swimmer's joint angles, see Fig. \ref{fig:models}. Using Lagrange's formulation, the equations of motion are given in matrix form as: 
\begin{equation}
\vecH(\vecq)\ddot{\vecq}+\vecB(\dot{\vecq},\vecq)=\vecF_\text{h}+\vecQ
\label{eq:EOMMatrix}
\end{equation}
where $\vecH$ is the swimmer's inertia matrix, $\vecB$ contains velocity-dependent terms, $\vecF_\text{h}$ is a vector of hydrodynamic forces applied by the fluid, and $\vecQ(t)=[0,0,0,\tau_1(t),\ldots,\tau_{N-1}(t)]^\text{T}$ contains generalized forces induced by the joints' torques. The inertia matrix $\vecH$ is related to the swimmer's kinetic energy $T$ through the relation $T=\frac{1}{2}\dot{\vecq}^\text{T}\vecH(\vecq)\dot\vecq$. This matrix can also be written explicitly as 
\begin{equation}
\vecH=\sum\limits_{i=1}^{N}{\vecJ_{i}^{\text{T}}(\vecq){{\vecM}_{i}}{{\vecJ}_{i}}(\vecq)}, \text{where} \,\, {{\vecM}_{i}}=\left[ \begin{matrix}
   {{m}_{i}} & 0 & 0  \\
   0 & {{m}_{i}} & 0  \\
   0 & 0 & {{I}_{i}}  \\
\end{matrix} \right].
\label{eq:inertiamatrix}
\end{equation}
The Jacobian matrices $\vecJ_i$ in (\ref{eq:inertiamatrix}) satisfy the kinematic relations $\vecv_i=\vecJ_i\dot{\vecq},$ where $\vecv_i=[\dot x'_i,\dot y'_i,\omega_i]^T$ is the linear and angular velocity of the $i$th link expressed in a frame $\mathcal{F}_i$ attached to its principal axes.
Using the ``perfect fluid'' model \cite{kanso2005locomotion,melli2006motion}, it is assumed that the fluid is governed by irrotational potential flow where viscous drag effects are neglected \cite{lamb1945hydrodynamics}. For simplicity, we follow \cite{hatton2013geometric} and neglect also the hydrodynamic interaction between the links. This implies that the hydrodynamic force acting on the $i$th link is decoupled from all other links, and satisfies
\begin{equation}
{\vecF_{i}}=\vecM'_i\veca_i,\, \text{where} \,\, \vecM'_i=\pi \rho \left[ \begin{matrix}
   b_{i}^{2} & 0 & 0  \\
   0 & a_{i}^{2} & 0  \\
   0 & 0 & \frac{1}{8}(a_{i}^{2}-b_{i}^{2})  \\
\end{matrix} \right],
\label{eq:added_mass}
\end{equation}
and $\veca_i$ is the linear and angular acceleration of the $i$th link expressed in the frame $\mathcal{F}_i$. The matrix $\vecM'_i$ in (\ref{eq:added_mass}) is the \textit{added mass tensor} of an ellipse-shaped body \cite{kanso2005locomotion,lamb1945hydrodynamics}, which is related to the momentum of the fluid that is displaced by the accelerating link. The relation (\ref{eq:added_mass}) enables elimination of the hydrodynamic forces $\vecF_\text{h}$ from (\ref{eq:EOMMatrix}) and replacing them by an addition to the system's kinetic energy and matrix of inertia, as: 
\begin{equation}
\begin{array}{l}
T=\frac{1}{2} \dot{\vecq}^{T}\tilde{\vecH}(\vecq)\dot{\vecq}=\frac{1}{2}\left( 
	\begin{matrix}
	   \vecv_b  \\
	   \dot{\vecq}_s  \\
	\end{matrix}
 \right)^\text{T} \left[
	 \begin{matrix}
		\vecM_{bb} & \vecM_{bs}\\
		\vecM_{bs}^T & \vecM_{ss}
	\end{matrix}
\right]
\left( 
	\begin{matrix}
	   \vecv_b  \\
	   \dot{\vecq}_s  \\
	\end{matrix}
 \right)\\
\text{where} \: \tilde{\vecH}(\vecq)=\sum\limits_{i=1}^{N}{\vecJ_i^{\text{T}}(\vecq)[{{\vecM}_{i}}+\vecM'_i]{{\vecJ}_{i}}(\vecq)} 
\end{array}
\label{eq:kinematic_energy}
\end{equation}
and $\vecv_b=[\dot x',\dot y',\omega_b]$ is the linear and angular velocity of the body frame $\mathcal{F}_b$ expressed in the frame $\mathcal{F}_b$. This body-fixed velocity is related to the swimmer's absolute body velocity via the kinematic equation
\begin{equation}
\dot \vecq_b=\vecR(\beta)\vecv_b, \:\: \text{where} \:\:\: \vecR(\beta)\!=\!
\left[\begin{matrix}
\cos \beta & -\sin \beta & 0\\
\sin \beta& \cos \beta & 0\\
0 & 0 & 1\\
\end{matrix}\right]
\label{eq:rotation}
\end{equation}
 The matrices $\vecM_{bb}, \vecM_{bs}$, and $\vecM_{ss}$ in (\ref{eq:kinematic_energy}), which depend only on the shape variables $\vecq_s$, are sub-blocks of $\tilde{\vecH}(\vecq)$ expressed in the frame $\mathcal{F}_b$ by substituting $\beta=0$. Note that the use of body-frame velocities $\vecv_b$ in (\ref{eq:kinematic_energy}) is possible due to the assumption of unbounded fluid domain that induces invariance of the dynamics with respect to rigid-body transformations (also known as  \textit{gauge symmetry} \cite{kelly1995geometric}). A well-known observation \cite{hatton2013geometric,ostrowski1998geometric} is that this invariance induces conservation of generalized momentum variables, formulated as:
\begin{equation}
\frac{d}{dt}\left(\vecM_{bb}(\vecq_s)\vecv_b+\vecM_{bs}(\vecq_s)\dot{\vecq}_s\right)=0
\end{equation}
Starting from rest $(\vecv_b=\dot{\vecq}_s=0)$ gives the relation between body velocity and shape changes, as:
\begin{equation}
\vecv_b=-\vecM_{bb}(\vecq_s)^{-1}\vecM_{bs}(\vecq_s)\dot{\vecq}_s=\vecA(\vecq_s)\dot{\vecq}_s
\label{eq:connections}
\end{equation}
Thus, the equation of motion (\ref{eq:EOMMatrix}) is reduced into a first-order system, augmented by the kinematic relation (\ref{eq:rotation}). Time-invariance of equation (\ref{eq:connections}) (also known as the system's \textit{connections } \cite{ostrowski1998geometric,kelly1995geometric}) implies that under a periodic input of shape changes, the net motion over a period depends only on the trajectory $\vecq_s(t)$ (i.e. \textit{gait}) and not on the time-rate of the motion.

\section{Asymptotic analysis of 3-link swimmer}

In this section we derive the leading-order expression and next order correction for the displacement of a 3-link swimmer over one period of harmonic inputs. First, we define some non-dimensional constants describing the swimmer's geometry. The ratio between the links' principal radii is denoted by a uniform $\alpha=b_i/a_i$ and the links' length ratio by $\eta=2l_0/l$ where $l$ is the full length of the swimmer $l=2(l_0+l_1+l_2)$. For simplicity, we assume that there is no spacing between the links, i.e. $a_i=l_i$.
The joint angles are given by $\theta_i=\varepsilon s_i(t)$, where $\varepsilon$ is the stroke amplitude and $s_i(t)$ is the unscaled gait trajectory given by:
\begin{equation}
s_1(t)= -\cos (t-\varphi/2), \: s_2(t)= \cos (t+\varphi/2)
\label{eq:unscaled_input}
\end{equation} 
with $t \in [0, 2\pi]$.
Equation (\ref{eq:connections}) now becomes:
\begin{equation}
\vecv_b=\vecA(\varepsilon,t)\varepsilon\dot{\vecs},
\end{equation}
where $\vecs=[s_1, s_2]^T$. This equation can be expanded as 
\begin{equation}
\vecv_b=\left(\vecA(0,t)\!+\!\varepsilon\!\left.\frac{\partial\vecA(\varepsilon,t)}{\partial\varepsilon}\right|_0\!+\!\varepsilon^2\frac{1}{2!}\!\!\left.\frac{\partial^2\vecA(\varepsilon,t)}{\partial\varepsilon^2}\right|_0\!+\!\ldots\right)\varepsilon\dot\vecs
\label{eq:V_expansion}
\end{equation}
Where all derivatives in (\ref{eq:V_expansion}) are evaluated at $\varepsilon=0$.
This gives the expansion of body-fixed velocities as:
\begin{equation}
\vecv_b(t)=\varepsilon\vecv_b^{(1)}+\varepsilon^2\vecv_b^{(2)}+\ldots
\label{eq:expansion}
\end{equation}
While the body position $x(t),y(t)$ cannot be directly integrated from the body fixed velocities $\dot x',\dot y'$, the orientation angle $\beta$ can be integrated from the expansion of $\omega_b(t)$ in (\ref{eq:expansion}) as $\beta(t)=\varepsilon\beta^{(1)}+\varepsilon^2\beta^{(2)}+\ldots$.
Next, we expand the rotation matrix $\vecR$ in (\ref{eq:rotation}) as:
\begin{align}
\vecR(\beta)&=\vecI+\beta
\left[\begin{matrix}
0 & -1 & 0\\
1 & 0 & 0\\
0 & 0 & 0\\
\end{matrix}\right]
+\beta^2
\left[\begin{matrix}
-1 & 0 & 0\\
0 & -1 & 0\\
0 & 0 & 0\\
\end{matrix}\right]+\ldots\nonumber\\&=
\vecI+\varepsilon\vecR^{(1)}+
\varepsilon^2\vecR^{(2)}+\ldots
\label{eq:expan_R}
\end{align}
Where $\vecI$ is the $3\times 3$ identity matrix.
Substituting the expansions for $\vecR$ in (\ref{eq:expan_R}) and for $\vecv_b$ in (\ref{eq:connections}) into (\ref{eq:rotation}) and rearranging into power series in $\varepsilon$, we obtain an expansion for $x(t)$ and $y(t)$. 
Due to symmetries of the gait in (\ref{eq:unscaled_input}), it can be shown that the net displacement in $y$ direction vanishes \cite{gutman2016symmetries,wiezel2016optimization}. The motion in $x$ direction can be obtained from integration over the period time:
\begin{equation}
X=\int_0^T \dot x(t) dt,
\end{equation}
which gives the following expansion:
\begin{align}
X&=\varepsilon^2 X^{(2)}+\varepsilon^4 X^{(4)}+O(\varepsilon^6)\label{eq:X_expansion}\\
\text{where,}\nonumber\\
X^{(2)}&=f_1(\eta)\sin \varphi > 0\nonumber\\
X^{(4)}&=f_2(\eta)\sin \varphi+f_3(\eta)\sin 2\varphi\nonumber
\end{align}
The functions $f_1(\eta)$, $f_2(\eta)$ and $f_3(\eta)$ depend on the links' aspect ratio $\alpha$ in a very cumbersome way. For concreteness, we choose $\alpha=0.5$ which is close to that of the experimental prototypes and gives much simpler expressions. The functions $f_1(\eta)$, $f_2(\eta)$ and $f_3(\eta)$ for $\alpha=0.5$ are given in Table \ref{tab:pol_eta}.
\begin{table}[!b]
\caption{Expressions from equation (\ref{eq:X_expansion}) for $\alpha=0.5$}
\resizebox{\textwidth}{!}{
  \begin{tabular}{|ll|}
  \hline &\\
  $f_1(\eta)=\!\dfrac{\pi l \eta{\left(\eta \! -\! 1\right)}^5\! \left(78 {\eta}^3 \!+\! 511 {\eta}^2 \!+\! 114 \eta \!+\! 29\right)}{4 {\left(3 {\eta}^2 \!-\! 2 \eta \!+\! 1\right)}^2 P_1(\eta)}$&$f_2(\eta)=\dfrac{ -\pi l \eta {\left(\eta - 1\right)}^5P_2(\eta)}{{64\left(3 {\eta}^2 - 2 \eta + 1\right)}^4 {P_1(\eta)}^3}$\\
 & \\
$f_3(\eta)=\dfrac{ \pi l \eta {\left(\eta - 1\right)}^5P_3(\eta)}{128{\left(3 {\eta}^2 - 2 \eta + 1\right)}^4 {P_1(\eta)}^3}$   & $P_1(\eta)=\left( \!-\! 333 {\eta}^4 \!+\! 196 {\eta}^3 \!+\! 170 {\eta}^2 \!+\! 116 \eta \!-\! 221\right)$\\
&\\
\multicolumn{2}{|l|}{$P_2(\eta)=3119734251 {\eta}^{15} \!-\! 3070539495 {\eta}^{14} \!-\! 11677468041 {\eta}^{13} \!+\! 25870509185 {\eta}^{12} \!-\! 19032800901 {\eta}^{11}\!+$}\\  
\multicolumn{2}{|l|}{$\qquad \quad \; \; \; \!5503302973 {\eta}^{10} \!-\! 4437032321 {\eta}^9\!+\!  
9942070757 {\eta}^8 \!-\! 10156151831 {\eta}^7 \!+\! 4574010219 {\eta}^6 +$}\\ 
\multicolumn{2}{|l|}{$\qquad \quad \; \; \; 973272381 {\eta}^5 \!-\! 2844406429 {\eta}^4 \!+\! 1987499057 {\eta}^3 \!-\! 771302273 {\eta}^2 \!+\! 189030989 \eta \!-\! 17941001$}\\
&\\
\multicolumn{2}{|l|}{$P_3(\eta)=2764445895 {\eta}^{15} - 8663576859 {\eta}^{14} + 9431287407 {\eta}^{13} - 4965255883 {\eta}^{12} + 6399003543 {\eta}^{11} - $}\\ 
\multicolumn{2}{|l|}{$\qquad \quad \; \; \; 12206561479 {\eta}^{10} + 12080221351 {\eta}^9 -
6386459751 {\eta}^8 + 1644304573 {\eta}^7 - 399495473 {\eta}^6 + $}\\
\multicolumn{2}{|l|}{$\qquad \quad \; \; \; 1027495461 {\eta}^5 - 1126099745 {\eta}^4 + 434875109 {\eta}^3 + 17293779 {\eta}^2 - 60307771 \eta + 20773779$}\\
\hline
\end{tabular}}
\label{tab:pol_eta}
\end{table}

For a phase difference of $\varphi>1[rad]$, $X^{(4)}$ is negative, and thus for large amplitude $\varepsilon$, the swimming direction is reversed. Moreover, there exists an optimal amplitude $\varepsilon^*$ that maximizes $X$, which is approximated from (\ref{eq:X_expansion}) as  $\varepsilon^*=\sqrt{ |X^{(4)}|/2X^{(2)}}$.
Next, we consider the influence of the phase difference $\varphi$ on the displacement $X$ for a given amplitude $\varepsilon$. From (\ref{eq:X_expansion}), it is obvious that $X$ vanishes for $\varphi=\{0,\pi\}$. This is because in these cases the shape change is time-reversible \cite{gutman2016symmetries,hatton2013geometric}. Moreover, there exists an intermediate value of optimal phase $\varphi^*$ that achieves maximal displacement. Considering only the leading-order term $X^{(2)}$ in (\ref{eq:X_expansion}) gives optimal phase of $\varphi^*=\pi/2$, but the next order term adds a correction to this optimal value. From (\ref{eq:X_expansion}), the optimal phase difference can be obtained as:
\begin{align}
\varphi^*&=\cos^{-1} \left[\left(-D_1 \pm \sqrt{D_1^2+16D_2^2}\right)/4D_2\right]\\
\text{where,}\nonumber\\
D_1&=\varepsilon^2f_1(\eta)+\varepsilon^4f_2(\eta)\text{ and }D_2=\varepsilon^4f_3(\eta)\nonumber
\end{align}

Additionally, we consider optimization with respect to the length ratio $\eta$ for a fixed total length $l$ of the swimmer. It can clearly be seen from (\ref{eq:X_expansion}) and Table \ref{tab:pol_eta} that for $\eta=0$ and $\eta=1$ the displacement $X$ vanishes, since one or two links of the swimmer have zero length.
Using only the leading-order expression $X^{(2)}$ in (\ref{eq:X_expansion}), the optimum of the polynomial $f_1(\eta)$ is numerically calculated as $\eta^*=0.3546$, indicating that the three links should be of nearly equal lengths. This result reveals a significant distinction from Purcell's 3-link microswimmer in a viscous fluid, whose optimal link ratio is $\eta\approx 0.25$, so that $l_1=l_2\approx 1.5l_0$.

Another possible manoeuvre of the swimmer is moderate turning obtained by performing small-amplitude oscillations about a constant angle $\gamma$ so that the joint angles are $\theta_1(t)=-\gamma-\varepsilon \cos (t-\varphi/2)$, $\theta_2(t)=\gamma+\varepsilon \cos (t+\varphi/2))$. The leading-order terms for the displacement $X$ and the net rotation $\Delta\beta$ under this actuation with $\eta=1/3$ and $\alpha=0.5$ are:
\begin{align}&\resizebox{0.48\textwidth}{!}{$
X^{(2)}=\frac{\pi l \sin\!(\varphi)\left( -\! 125184 C^5\! -\! 355448 C^4 \!-\! 32802 C^3 \!+\! 743779 C^2 \!+\! 848034 C \!+\! 309286\right)}{9 \left(C^2 \!+\! 2\right) {\left(262 C^2 \!+\! 768 C \!+\! 593\right)}^2}$}\label{eq:X_dis_gamma}\\&
\resizebox{0.39\textwidth}{!}{$
\Delta\beta^{(2)}=\frac{128 \pi  \sin\!(\varphi) S \left( -\! 652 C^4 \!-\! 1437 C^3 \!+\! 455 C^2 \!+\! 3339 C \!+\! 2534\right)}{\left(C^2 \!+\! 2\right) {\left(262 C^2 \!+\! 768 C \!+\! 593\right)}^2}$}
\label{eq:deltabeta}
\end{align}
where $C=\cos(\gamma)$ and $S=\sin(\gamma)$. The net displacement in the $y$ direction is only of order $O(\varepsilon^4)$.
Eqs. (\ref{eq:X_dis_gamma})-(\ref{eq:deltabeta}) show that in addition to the displacement in the $x$ direction, the swimmer has net rotation $\Delta\beta$ over a period. This allows the swimmer to perform an arclike motion.  
The curvature of the resulting trajectory of the swimmer $\kappa=\frac{\Delta\beta}{X}$ for a small offset angle $\gamma$ is $\kappa=3.52\gamma/l$. Animations of the simulated motion of the swimmer under this actuation can be found in the multimedia extension.

\begin{figure}[!b]
      \centering
      \begin{subfigure}[h]{0.3\textwidth}
      \centering
      \includegraphics[width=\textwidth]{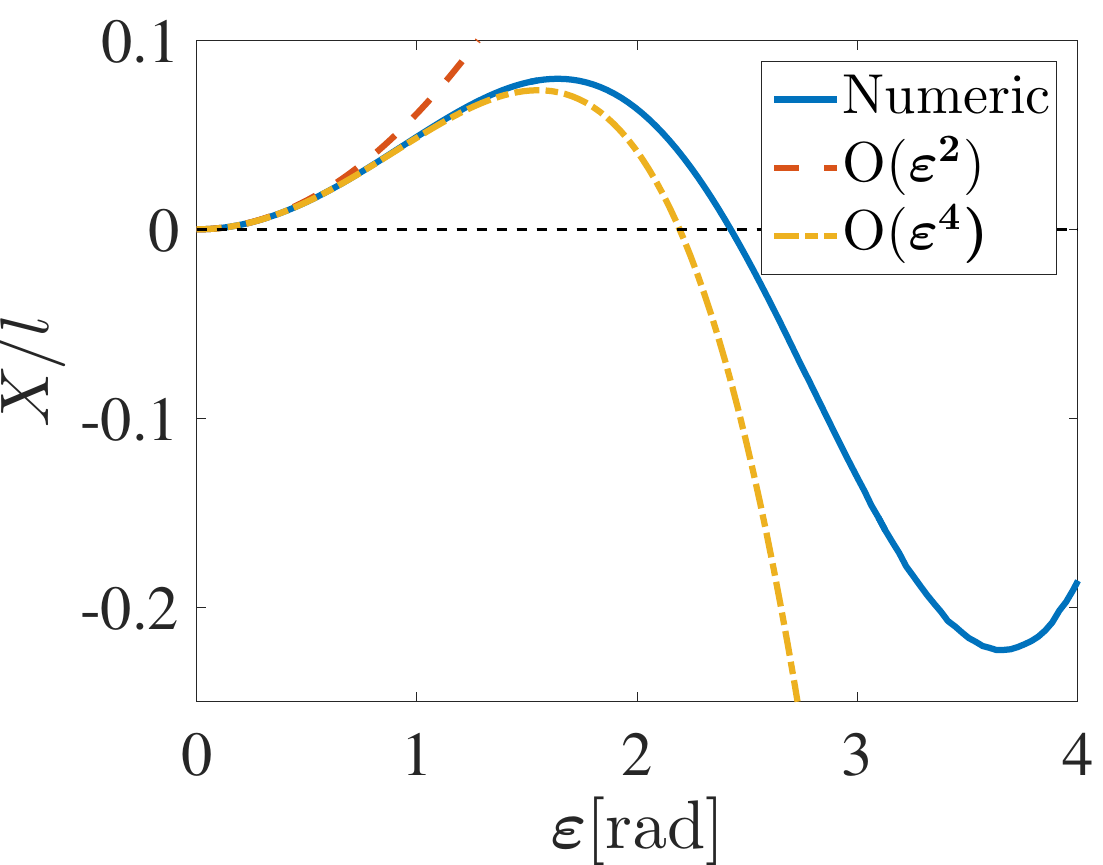}
      \caption{}
      \label{fig:assym_X_vs_eps}
      \end{subfigure}
      \begin{subfigure}[h]{0.3\textwidth}
      \centering
      \includegraphics[width=\textwidth]{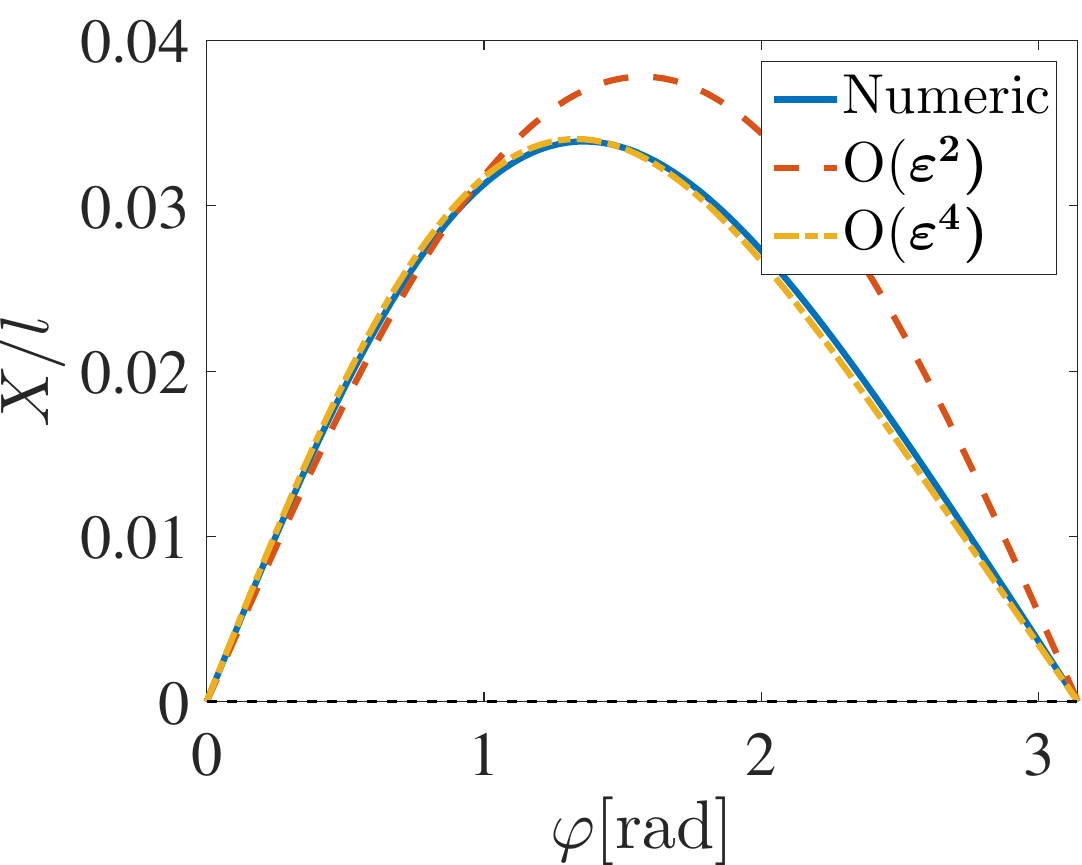}
      \caption{}
      \label{fig:assym_X_vs_p}
      \end{subfigure}
      \begin{subfigure}[h]{0.3\textwidth}
      \centering
      \includegraphics[width=\textwidth]{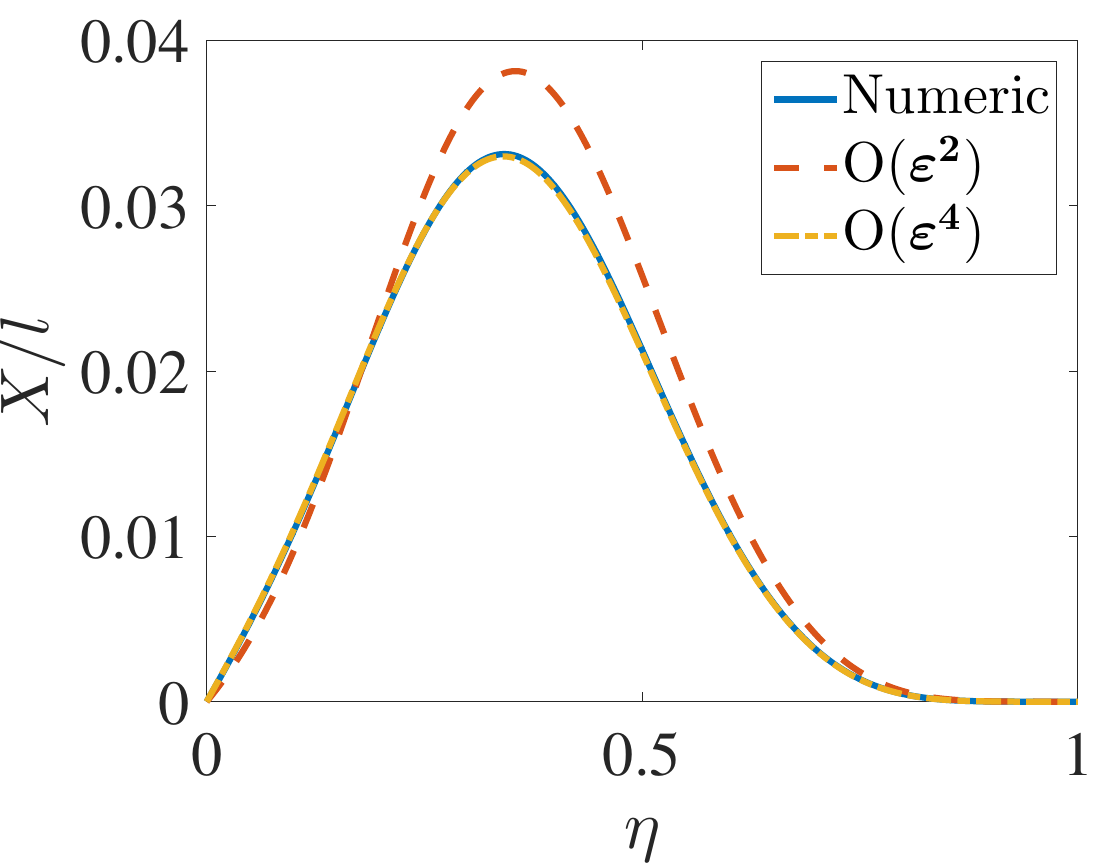}
      \caption{}
      \label{fig:assym_X_vs_eta}
      \end{subfigure}
      \caption{Numerical simulation and asymptotic approximations for the three link swimmer. (a) $X$ vs $\varepsilon$ for $\varphi=\pi/2$, $\eta=1/3$, (b) $X$ vs $\varphi$ for $\varepsilon=\pi/4$, $\eta=1/3$, (c) $X$ vs $\eta$ for $\varepsilon=\pi/4$, $\varphi=\pi/2$.}
      \label{fig:sim_vs_assym}
   \end{figure}

\section{Numerical simulations and gaits}

We now present the results of numerical simulations of the motion of a 3-link swimmer and compare to the asymptotic approximation.
Additionally, we numerically obtain the optimal combination of gait amplitude and phase difference for both 3-link and 5-link swimmers.
In Fig. \ref{fig:sim_vs_assym}, the solid lines represent the numerical calculation, the dashed lines represent the results using only the leading-order approximation and the dash-dotted lines are the results with the next order correction.
Numerical integration of the dynamic equation of motion (\ref{eq:connections}) has been performed using adaptive Runge-Kutta procedure $\mathbf{ode45}$ in Matlab.
Fig. \ref{fig:assym_X_vs_eps} shows the $X$ displacement over a period for varying amplitudes and a phase difference of $\varphi=\pi/2$. It can be seen that for large amplitudes the swimming direction is reversed. Obviously, the reversal cannot be seen in the leading-order results which are quadratic in $\varepsilon$ and monotonic.
Nevertheless, including the next order term $X^{(4)}$ does show this behaviour and has an optimal amplitude. The optimal amplitude using the numerical calculation is $\varepsilon^*=1.65[rad]$ with a normalized displacement of $X=0.079l$ and through the asymptotic approximation $\varepsilon^*=1.55[rad]$ with a displacement of $X=0.074l$. For larger amplitudes of $\varepsilon>\pi$, it is shown in Fig. \ref{fig:assym_X_vs_eps} that there exists another optimum with negative displacement that has even larger absolute value. However, in these large amplitudes the swimmer's links will collide and thus, this result is regarded as infeasible.  
Fig. \ref{fig:assym_X_vs_p} shows the displacement as a function of the phase difference $\varphi$ with an amplitude of $\varepsilon=\pi/4$. For a phase difference of $\varphi=\{0,\pi\}$ the displacement is zero as expected from (\ref{eq:X_expansion}). The optimal phase that maximizes the displacement $X$ is $\varphi^*=1.36[rad]$ for the numerical calculation with a displacement of $X=0.034l$, while the asymptotic approximation gives an optimal phase of $\varphi^*=1.33$ with displacement of $X=0.034l$.
\setcounter{figure}{2}
\begin{figure}[!h]
	   \centering
	   \includegraphics[width=0.6\textwidth]{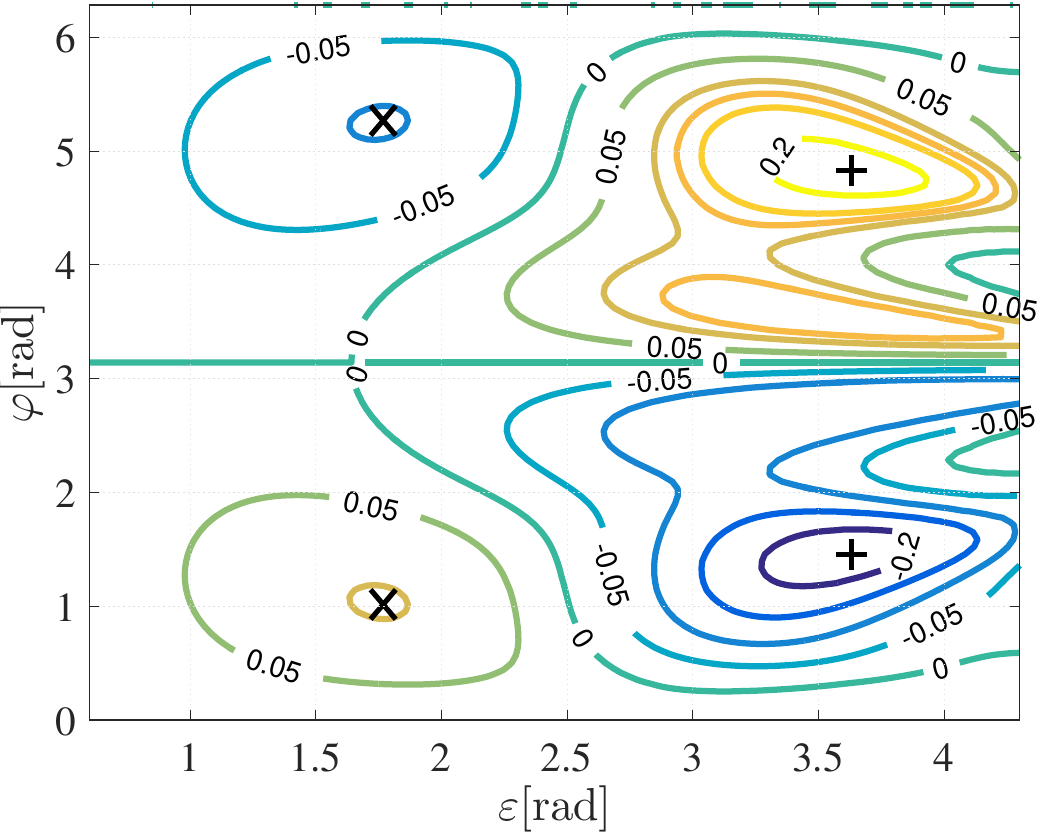}
	   \caption{Contour plot of net displacement $X$ of the 3-link swimmer as a function of amplitude $\varepsilon$ and phase $\varphi$. The points of maximal displacement at $|X|=0.102l$ are marked by `$\times$'. The `$+$' markers denote maximal distance $|X|=0.23l$ which is attained only for unphysical values of joint angle amplitudes.}
	   \label{fig:contour_3link}
   \end{figure}
   %
    \begin{figure}[!b]
	   \centering
	   \includegraphics[width=0.6\textwidth]{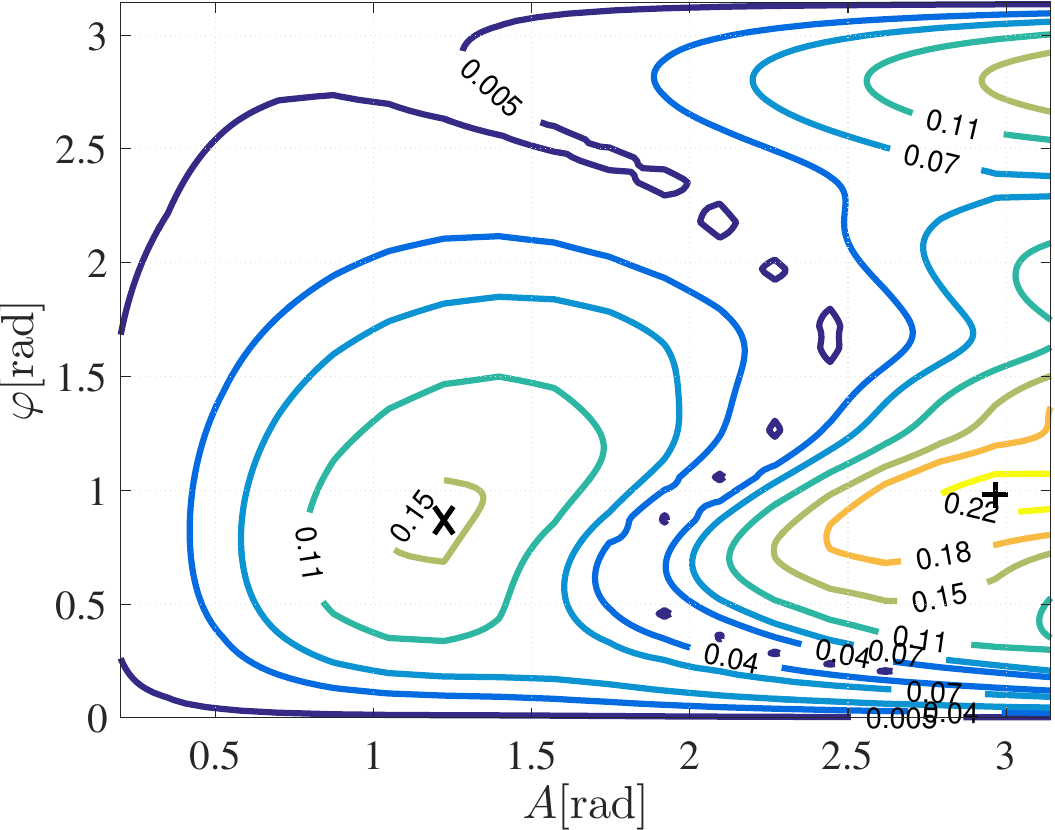}
	   \caption{Contour plot of net displacement $d$ of the 5-link swimmer as a function of amplitude $A$ and phase $\varphi$.}
	   \label{fig:contour_5link}
   \end{figure}
Fig. \ref{fig:assym_X_vs_eta} shows the displacement for a given gait ($\varepsilon=\pi/4$ and $\varphi=\pi/2$) with varying links' length ratio $\eta$. Both the numeric simulation and the fourth order approximation give a similar optimal ratio $\eta^*=0.34$ for the given gait, with a displacement of $X=0.033l$. The leading order as well gives a close approximation of the optimal ratio $\eta^*=0.35$, but slightly misses the displacement, with $X=0.038l$. 
Fig. \ref{fig:contour_3link} shows a contour plot of the displacement $X$ as a function of the amplitude $\varepsilon$ and phase difference $\varphi$ for the 3-link swimmer through numerical integration. The optimal combination of  amplitude and phase, marked by `$\times$' on the plot, is at $\varepsilon^*=1.74$, $\varphi^*=1.01$ with a displacement of $X=0.102l$. (This optimum cannot be captured by the asymptotic solution in (\ref{eq:X_expansion}) without considering the $O(\varepsilon^6)$ term). The additional global optima, marked by a `$+$' on the right edge of Fig. \ref{fig:contour_3link}, are in the range of large amplitudes which are not feasible due to inter-collision between links.

%
For the 5-link swimmer model, we performed simulations under harmonic inputs $\theta_k(t)=A\sin (\omega t + k\varphi)$ with identical links of $a_i=1$, $b_i=0.5$ and $l_i=1.1$.
Fig. \ref{fig:contour_5link} shows the contour plot of the displacement $X$ of the 5-link swimmer as a function of the amplitude $A$ and phase difference $\varphi$. The optimal combination of  amplitude and phase, marked by `$\times$', is $A^*=1.22[rad]$ and $\varphi^*=0.87[rad]$ with a displacement of $X=0.156l$. Another maximum, with a greater displacement, is marked by a `$+$'.
As before, this is not considered since the amplitude is greater than $\pi$ and collision between the links will occur before the swimmer reaches this point.
Motion animations of the simulated swimmers appear in the multimedia extension.

\section{Experimental results}

   \setcounter{figure}{4}
   \begin{figure}[!b]
	\centering
	\begin{subfigure}[h]{0.4\textwidth}
			\includegraphics[width=0.28\textwidth, angle=-90]{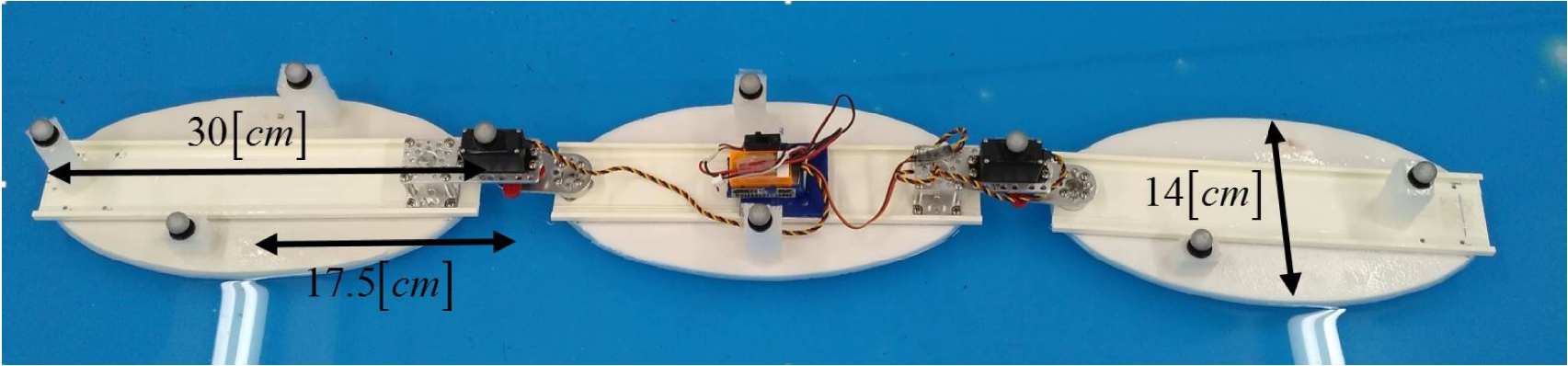}
			\caption{}
			\label{fig:robot_3link}
	    \end{subfigure}
	    \hspace{1.5cm}
	    \begin{subfigure}[h]{0.45\textwidth}
			\includegraphics[width=0.25\textwidth, angle=-90]{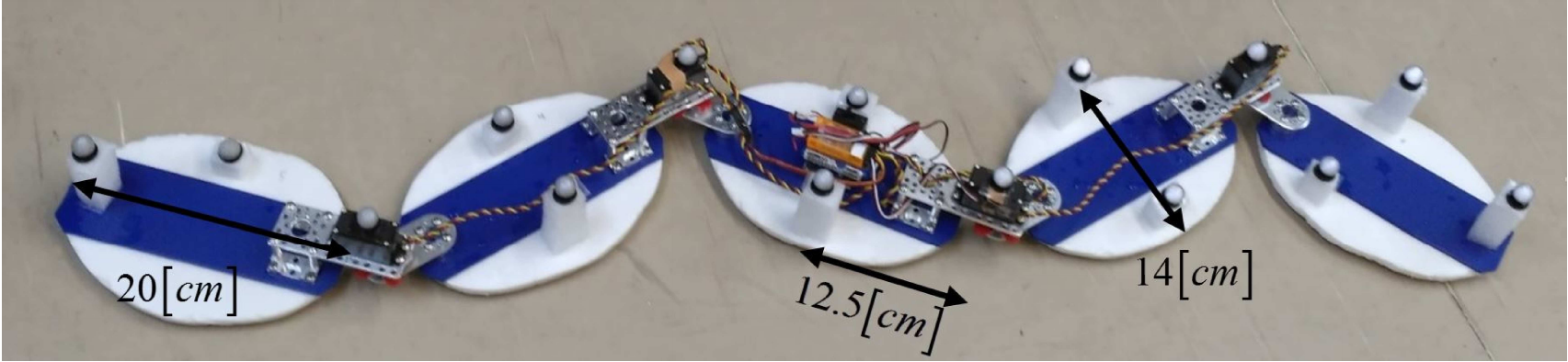}
			\caption{}
			\label{fig:robot_5link}
    \end{subfigure}
    \caption{Robotic prototypes. (a) 3-link robotic swimmer ($a_i=15$, $b_i=7$, $l_i=17.5$). (b) 5-link robotic swimmer.($a_i=10$, $b_i=7$, $l_i=12.5$, all dimensions in $cm$).}
\end{figure}

\setcounter{figure}{5}
\begin{figure}[!t]
	   \centering
	   \includegraphics[width=0.5\textwidth]{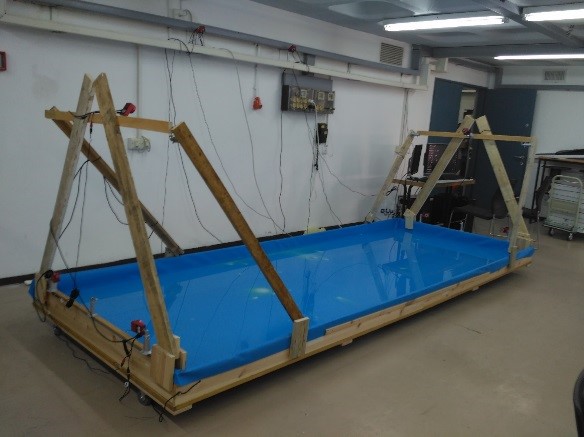}
	   \caption{Experimental setup. Water pool with \textit{Optitrack} cameras.}
	   \label{fig:pool}
   \end{figure}

We now present experimental results that have been obtained with untethered floating 3-link and 5-link swimming robots. Prototypes of these robots and their dimensions are shown in Figs. \ref{fig:robot_3link},\ref{fig:robot_5link}.
Their links were made of ellipse-shaped flotation foams of thickness 1$cm$ for the 3-link swimmer and 2$cm$ for the 5-link swimmer. The links were connected by joints which are actuated by servo motors (Hitec Multiplex HS-5685$MH$) that were mounted on top of the floating links. A single battery (2-cell 7.4$V$ Turnigy 2$s$ 500$mAh$ Lipo) for powering the motors and RF receiver (orangeRx R615X) were mounted on top of the middle link. Harmonic inputs for the joint angles as in (\ref{eq:input}) were fed from MATLAB interface to CRIO-Labview system, and then transmitted to the onboard RF receiver and servo motors, in order to track coordinated reference trajectories. The robots were located in a rectangular pool (length 401$cm$, width 151$cm$, height 18$cm$), which has been filled with water up to a level of 6$cm$ (Fig. \ref{fig:pool}). Three spherical reflective markers have been attached to each link, and the robots' motion was tracked by Optitrack system consisting of an array of eight infrared cameras. The spatial location of each link has been measured with sampling rate of 100$Hz$, and then processed in Motive tracking software. The resulting position vectors were smoothened by a moving average filter with 25-points window in order to extract the trajectories of robot's position and joint angles. 

\begin{figure}[!t]
	\centering
	\begin{subfigure}[h]{0.49\textwidth}
			\includegraphics[width=\textwidth]{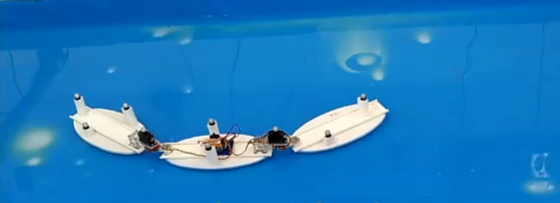}
			\caption*{$t=0$}
	    \end{subfigure}
	    \begin{subfigure}[h]{0.49\textwidth}
			\includegraphics[width=\textwidth]{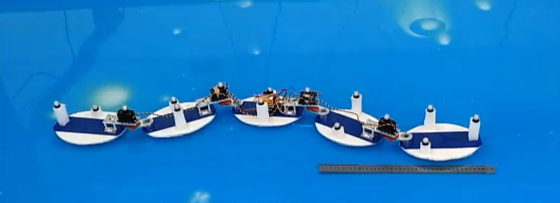}
			\caption*{$t=0$}
	    \end{subfigure}
	\begin{subfigure}[h]{0.49\textwidth}
			\includegraphics[width=\textwidth]{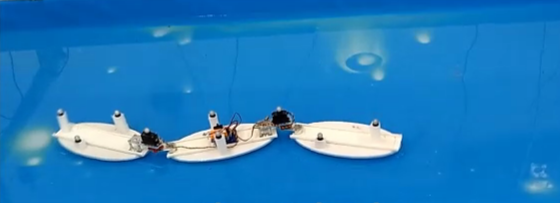}
			\caption*{$t=0.25T$}
    \end{subfigure}
    \begin{subfigure}[h]{0.49\textwidth}
			\includegraphics[width=\textwidth]{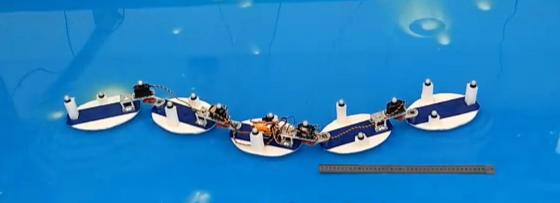}
			\caption*{$t=0.25T$}
	    \end{subfigure}
    \begin{subfigure}[h]{0.49\textwidth}
			\includegraphics[width=\textwidth]{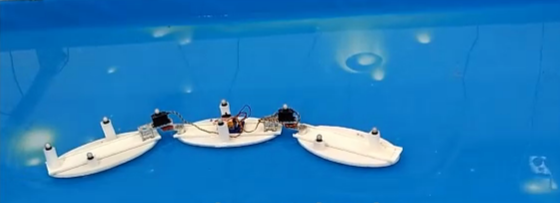}
			\caption*{$t=0.5T$}
    \end{subfigure}
    \begin{subfigure}[h]{0.49\textwidth}
			\includegraphics[width=\textwidth]{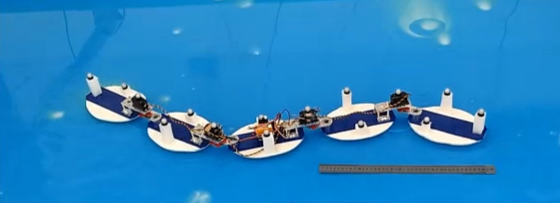}
			\caption*{$t=0.5T$}
	    \end{subfigure}
    \begin{subfigure}[h]{0.49\textwidth}
			\includegraphics[width=\textwidth]{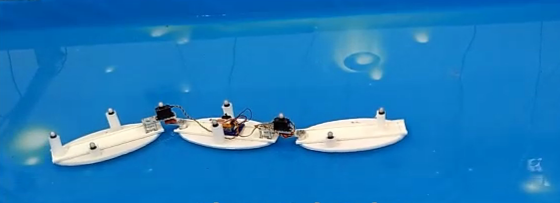}
			\caption*{$t=0.75T$}
    \end{subfigure}
    \begin{subfigure}[h]{0.49\textwidth}
			\includegraphics[width=\textwidth]{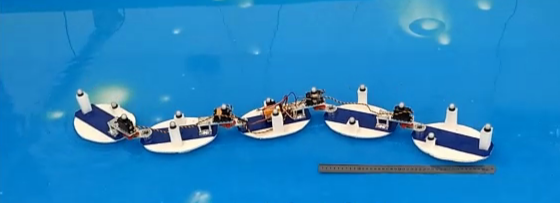}
			\caption*{$t=0.75T$}
	    \end{subfigure}
    \begin{subfigure}[h]{0.49\textwidth}
			\includegraphics[width=\textwidth]{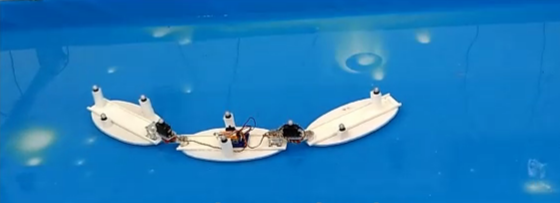}
			\caption*{$t=T$}
    \end{subfigure}
    \begin{subfigure}[h]{0.49\textwidth}
			\includegraphics[width=\textwidth]{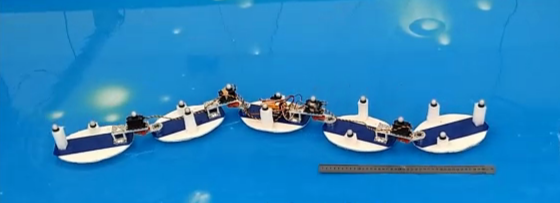}
			\caption*{$t=T$}
	    \end{subfigure}
    \caption{Motion snapshots of robotic prototypes of 3- and 5-link swimmers. $T=2\pi/\omega$ is the period time of the inputs.}
    \label{fig:time_snapshots}
\end{figure}

\begin{figure}[!t]
	\centering
	\begin{subfigure}[h]{0.32\textwidth}
		\centering
		\includegraphics[width=\textwidth]{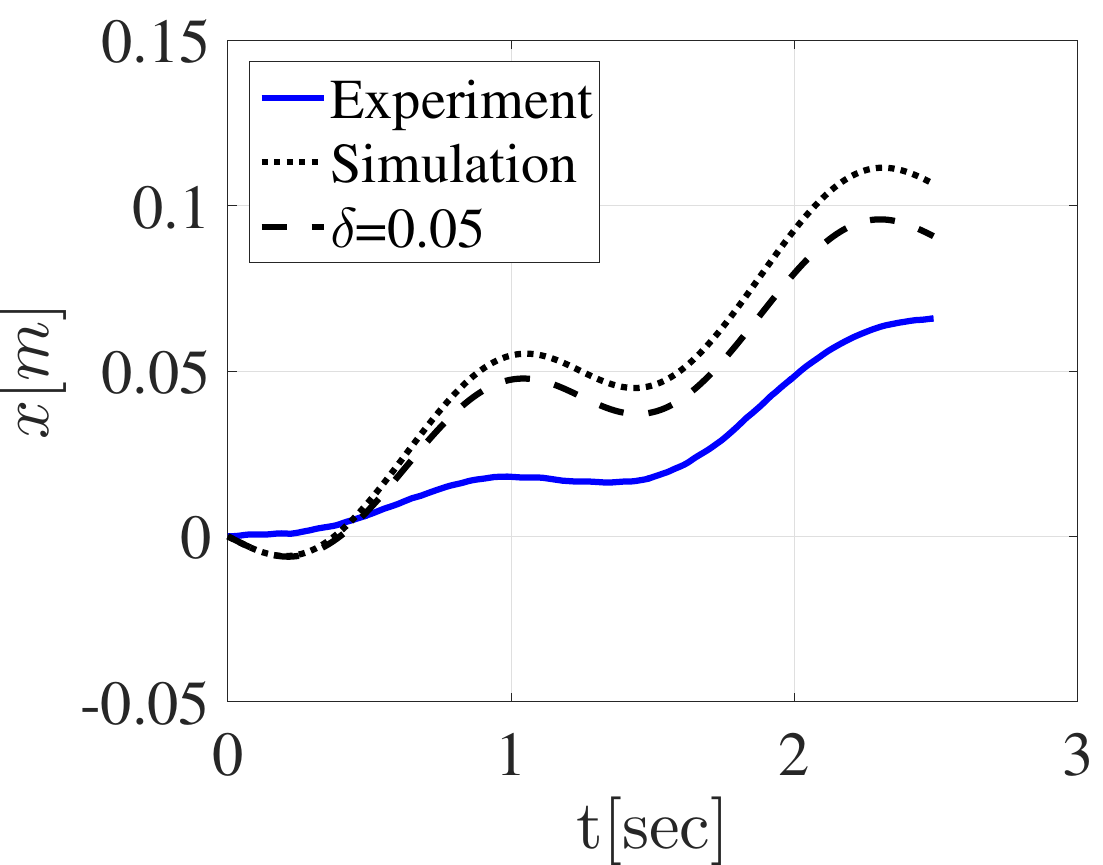}
		\caption{}
	    \label{fig:exp_x(t)3l}
    \end{subfigure}
	\begin{subfigure}[h]{0.32\textwidth}
		\centering
		\includegraphics[width=\textwidth]{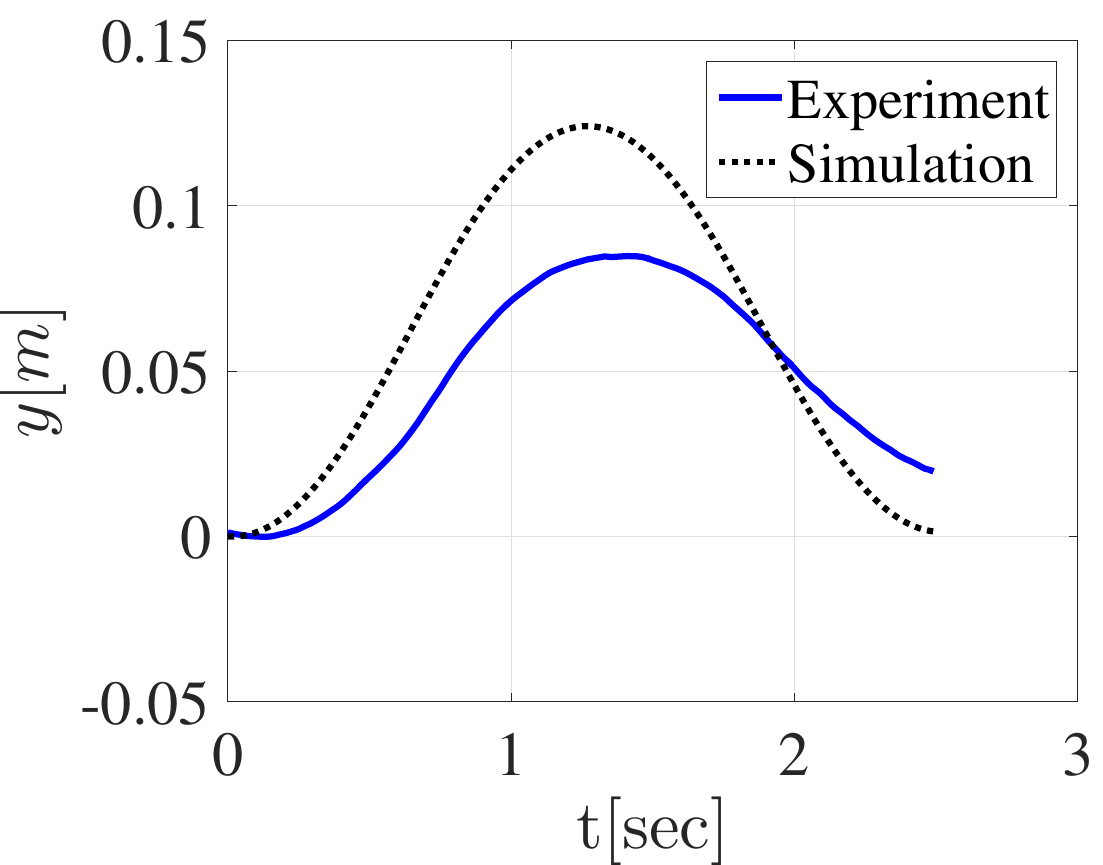}
		\caption{}
	    \label{fig:exp_y(t)3l}
    \end{subfigure}
    \begin{subfigure}[h]{0.32\textwidth}
		\centering
		\includegraphics[width=\textwidth]{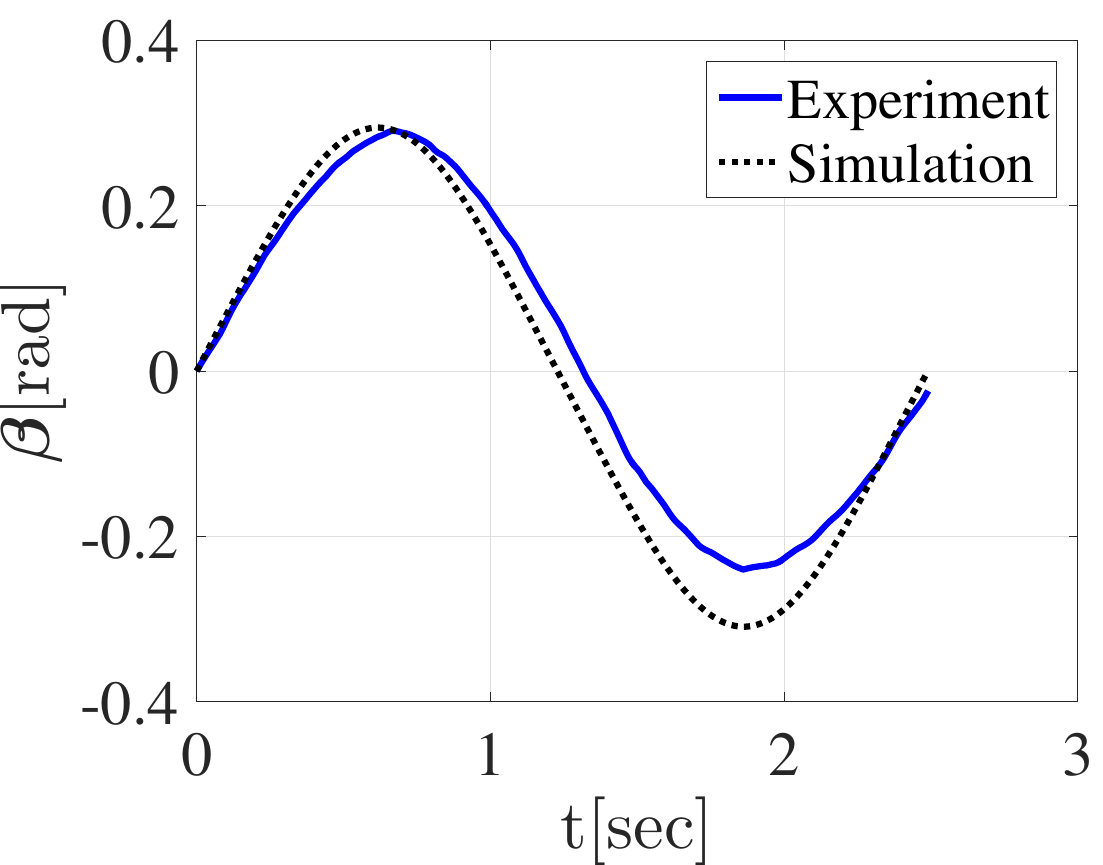}
		\caption{}
	    \label{fig:exp_beta(t)3l}
    \end{subfigure}
    \caption{Experimental results for 3-link swimmer: (a) $x(t)$, (b) $y(t)$, (c) $\beta (t)$.}
    \label{fig:exp3l_xyb}
\end{figure}

Motion experiments were conducted for both 3-link and 5-link swimmers under several input parameters, and the measured results have been compared to numerical simulations under the same joint kinematics as extracted from the measurements.
A video file that appears in the multimedia extension of this paper presents the experimental setup, motion animations of numerical simulations, as well as movies of representative swimming experiments. Fig.  \ref{fig:time_snapshots} shows motion snapshots of the 3-link and 5-link robotic swimmers, which are taken from the movie.
For the 3-link swimmer, Figs. \ref{fig:exp_x(t)3l},\ref{fig:exp_y(t)3l},\ref{fig:exp_beta(t)3l} show time plots of the body position $x(t)$,$y(t)$,$\beta(t)$, respectively, during a single period, under inputs as in (\ref{eq:unscaled_input}) with $\varepsilon=0.78[rad]$ and $\varphi=0.25[rad]$.
The solid lines denote the experimental measurements, while the dotted lines denote numerical simulations. It can be seen that the motions of lateral translation $y(t)$ and rotation $\beta(t)$ display reasonable agreement with numerical simulations, whereas the forward motion $x(t)$ is significantly overestimated by the simulations.
One obvious explanation to this difference is the fact that the model accounts for a \textit{fully submerged} robot while in reality, only a small portion of the ellipses is submerged and all masses of the motors, batteries and receiver contribute to the robot's inertia but not to the added mass effect which generates propulsion. This observation can be easily incorporated into the theoretical model by introducing a mass reduction coefficient $\delta$, which is the ratio between the submerged part of the link's mass to its total mass.
For our swimmer's mass and buoyancy parameters, this coefficient is estimated as $\delta=0.05$, and the dashed line in Fig. \ref{fig:exp_x(t)3l} denotes the simulated motion while considering this mass reduction, i.e. multiplying the added mass terms in (\ref{eq:kinematic_energy}) by $\delta$.
It can be seen that this gives a noticeable improvement in the quantitative agreement between experimental measurements and numerical simulations of $x(t)$. 

\begin{figure}[!t]
	\centering
	\begin{subfigure}[h]{0.4\textwidth}
		\centering
		\includegraphics[width=\textwidth]{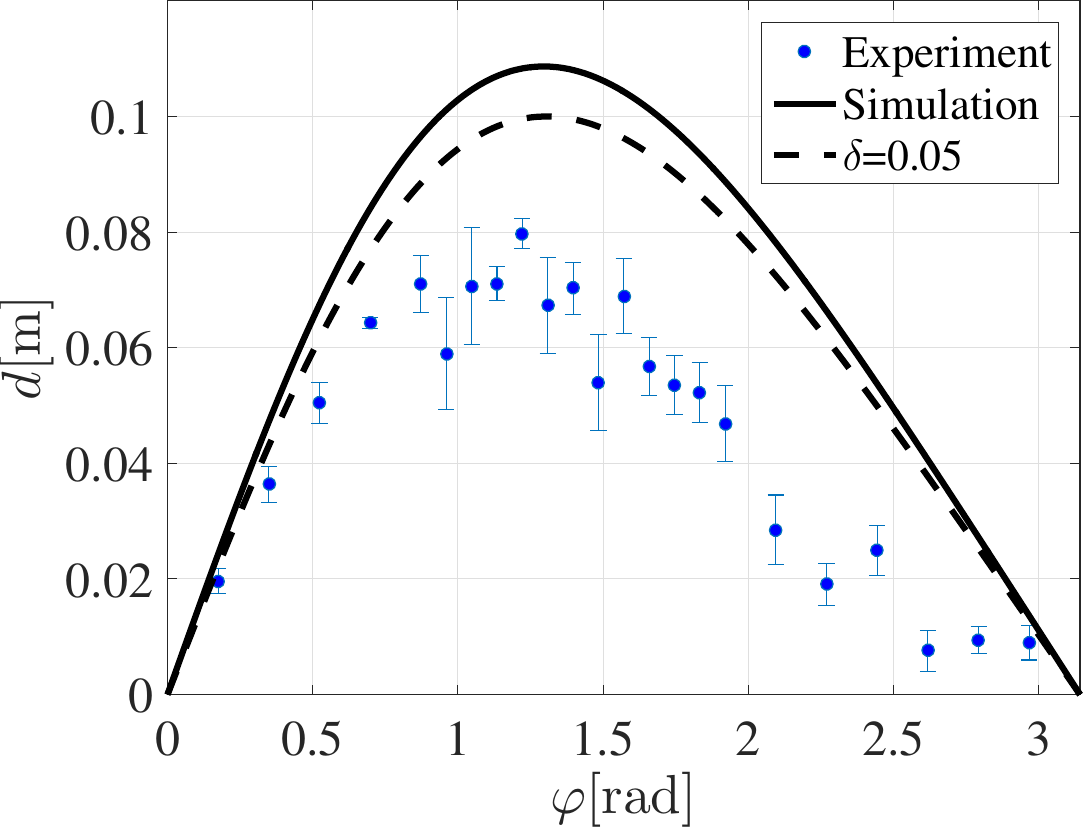}
		\caption{}
	    \label{fig:exp3l_X_vs_p}
    \end{subfigure}
	\begin{subfigure}[h]{0.4\textwidth}
		\centering
		\includegraphics[width=\textwidth]{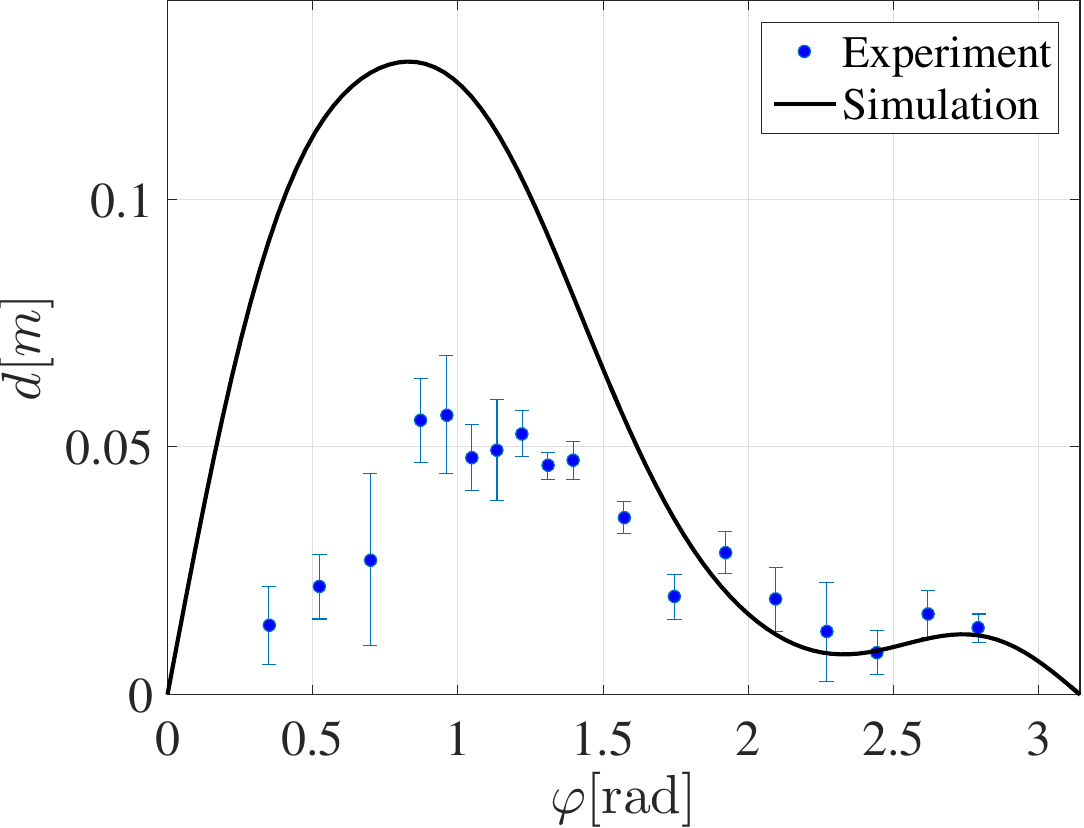}
		\caption{}
	    \label{fig:exp5l_X_vs_p}
    \end{subfigure}
    \caption{Experimental results: (a) 3-link swimmer - $X$ vs $\varphi$ , (b) 5-link swimmer - $d$ vs $\varphi$.}
\end{figure}
\begin{figure}[!t]
	\centering
	\begin{subfigure}[h]{0.32\textwidth}
		\centering
		\includegraphics[width=\textwidth]{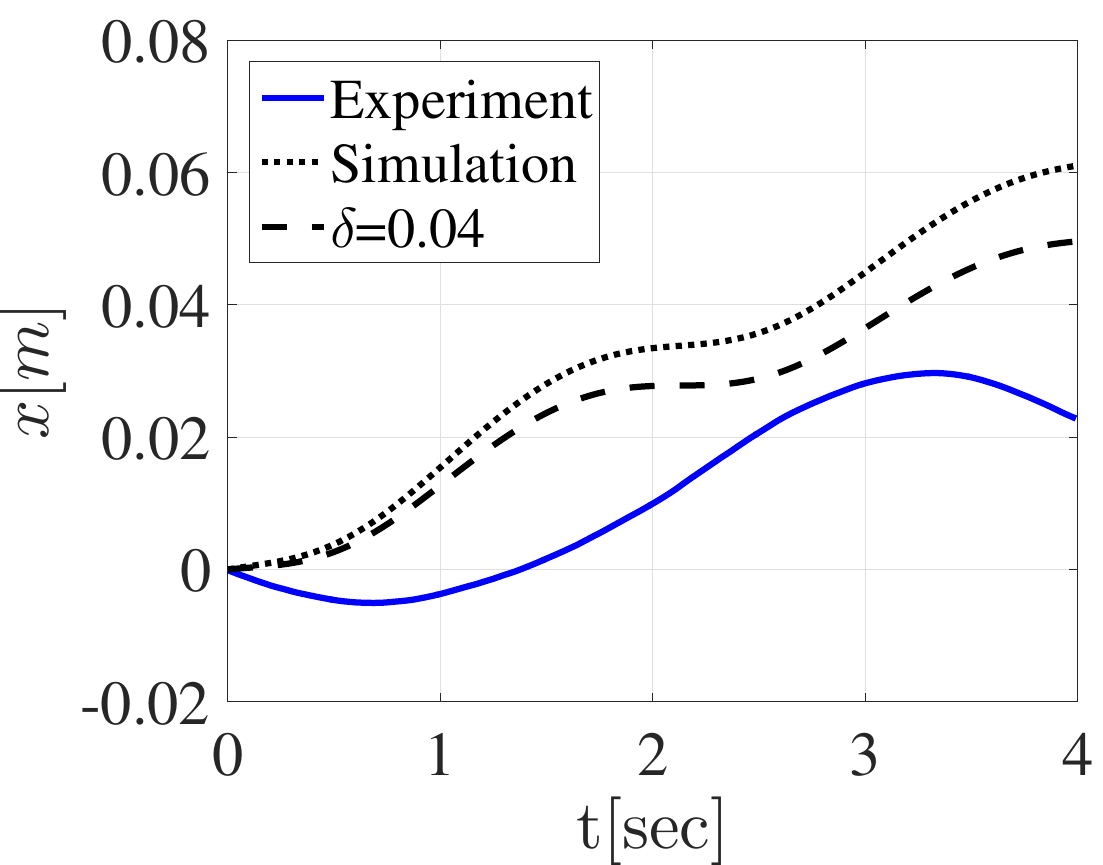}
		\caption{}
	    \label{fig:exp_x(t)5l}
    \end{subfigure}
	\begin{subfigure}[h]{0.32\textwidth}
		\centering
		\includegraphics[width=\textwidth]{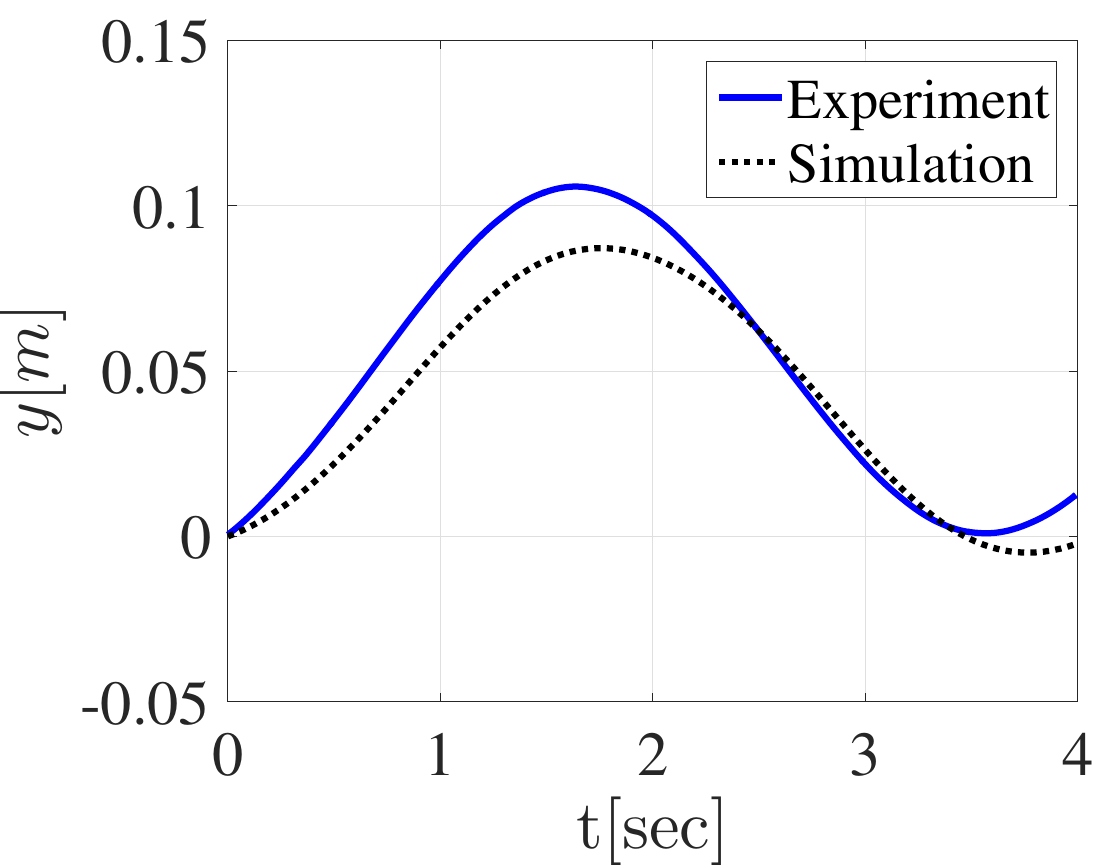}
		\caption{}
	    \label{fig:exp_y(t)5l}
    \end{subfigure}
    \begin{subfigure}[h]{0.32\textwidth}
		\centering
		\includegraphics[width=\textwidth]{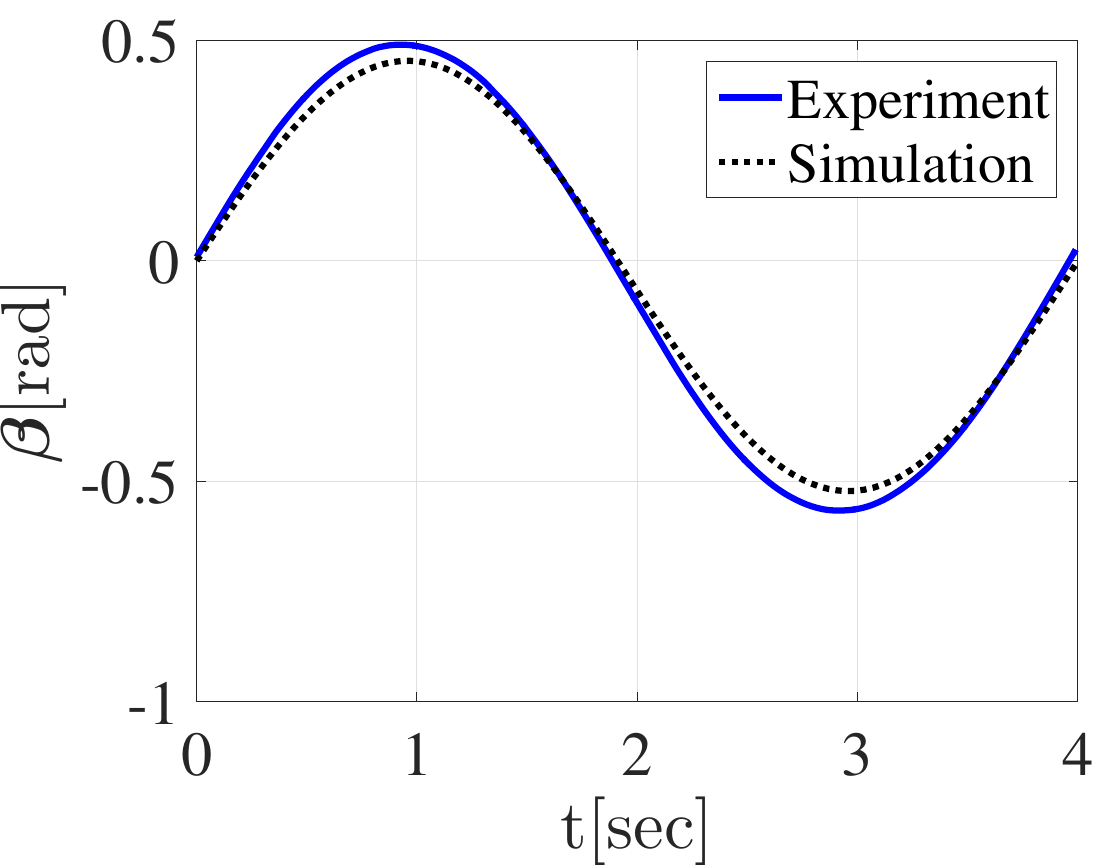}
		\caption{}
	    \label{fig:exp_beta(t)5l}
    \end{subfigure}
    \caption{Experimental results for 5-link swimmer: (a) $x(t)$, (b) $y(t)$, (c) $\beta(t)$.}
    \label{fig:exp_5l}
\end{figure}

Next, we conducted a series of experiments with inputs of the form (\ref{eq:unscaled_input}), where the amplitude was kept constant at $\varepsilon=0.78[rad]$ while the phase difference $\varphi$ between the two joint angles have been varied in $5$ degree increments.
Fig. \ref{fig:exp3l_X_vs_p} plots the forward displacement $X$ in a period as a function of the phase difference $\varphi$.
The circular markers denote experimental measurements which were averaged over 3 periods, where the error bars denote standard deviations.
The solid line denotes numerical simulations under the same inputs without mass reduction, while the dashed line denotes simulation results under mass reduction of $\delta=0.05$.
It can be seen that the experimental results corroborate the theoretical predictions of an optimal phase difference at $\varphi\approx1.3[rad]$ that achieves maximal displacement. 
Moreover, adding the mass reduction factor $\delta$ into the theoretical model improves the quantitative agreement with experimental measurements.
Similar experiments have been conducted for the five-link swimmer. Fig. \ref{fig:exp_5l} shows time plots of the body position $x(t)$,$y(t)$,$\beta(t)$, respectively, under inputs $\theta_k(t)=0.48\sin(0.5\pi t+k\varphi)[rad]$ for $k=1\ldots 4$ and phase difference $\varphi=-\pi/4[rad]$.
Fig. \ref{fig:exp5l_X_vs_p} plots the net swimming distance $d=\sqrt{\Delta x^2+\Delta y^2}$ as a function of the phase difference $\varphi$  between consecutive joint angles.

One can see a good qualitative agreement between experimental results and simulations of the theoretical model, which both capture similar behaviour of $x(t)$,$y(t)$,$\beta(t)$ during a cycle, and also show an optimal phase difference of $\varphi\approx0.83[rad]$ that achieves maximal displacement.
Nevertheless, the quantitative agreement between theory and experiment for this swimmer is weaker than that of the three-link swimmer.
Incorporating the effect of added mass reduction does not result in significant improvements (shown in Fig. \ref{fig:exp_x(t)5l} only).
This suggests that for the five-link swimmer, other unmodelled effects are more dominant, as discussed next.

%
%

\section{Discussion}
 
We now discuss the results and make some observations regarding the comparison between the experiments and the numerical simulations based on the theoretical model. It is important to note that the theoretical model of ``perfect fluid'' is highly simplistic and thus limited. It does not account for many realistic effects that are obviously present in the experimental prototypes, listed as follows. 
First, the model does not account for drag forces generated due to the fluid's viscosity \cite{morgansen2002trajectory,crespi2008online,porez2014improved}. It also ignores the effects of hydrodynamic interaction between the links \cite{kanso2005locomotion,melli2006motion}, and of vortex shedding that enhances propulsion \cite{tallapragada2015self,zhu2002three}.
These effects have been previously modelled by other works as mentioned above. Nevertheless, these more accurate models are significantly more complicated, and result in a major increase in computational resources and run-time complexity, while symmetries and time-invariance of the low-dimensional ``perfect fluid'' model are typically lost.
Second, the model assumes an unbounded fluid domain, while reflected waves from the pool's walls can have a significant effect on the robot's motion. This effect has been strongly observed for the five-link swimmer, whose larger total length ($120cm$) becomes comparable to the dimensions of the pool.
Third, as mentioned above, the experimental swimmer prototype floats while only small portion of the links is submerged in the fluid, whereas the theoretical model assumes that the entire swimmer is submerged, and thus ignores  the effects of surface tension at the water-air-swimmer interface.
Additionally, this work focuses on optimizing harmonic inputs only, whereas an important extension can be considering optimization of time-periodic input trajectories of any shape, as done in previous works for dynamic locomotion systems \cite{cortes2001optimal,ostrowski2000optimal} as well as quasistatic motion of multi-link micro-swimmers \cite{tam2007optimal,wiezel2016using,wiezel2018energy,ramasamy2017geometric}.
Finally, the theoretical model considers only planar horizontal (gravity-free) motion, while the real swimmer can undergo off-plane motion. In some experiments, the swimmer has displayed noticeable off-plane rocking motion similar to a gravity-dominated pendulum. These oscillations were particularly emphasized in cases of large joint angles and ``U-shaped'' configurations of the swimmer. This effect, combined with mechanical limitation on joint angles due to inter-link collisions, did not enable conducting experiments with large stroke amplitudes of the joint angles for corroborating the theoretical predictions of optimal amplitude. This task is left as a future challenge, that requires improved mechanical design of the swimmer.

\section{Conclusions}

In this paper, we have studied the inertia-dominated motion of multi-link swimmers under harmonic inputs of joint angles. We utilized the ``perfect fluid'' model that accounts for added mass effect and assumes ideal inviscid fluid, which enables reduction to a time-invariant first-order dynamical system. We conducted asymptotic analysis for the three-link swimmer, which gives closed-form approximate expressions for the swimmer's displacement, that enable obtaining optimal amplitude and phase shift for the joint angles, as well as optimal ratio of links' length. Next, we conducted motion experiments with three-link and five-link floating swimmers, and compared measurements from motion tracking system to numerical simulations under the theoretical model, while accounting for the reduction in added mass due to the swimmer's buoyancy. Very good agreement has been achieved for the three-link swimmer, while the results of the five-link swimmer agree only qualitatively. We discussed possible reasons for the discrepancies, mainly due to wall interactions and other unmodelled effects.
Future work will include optimization of general gait trajectories, either for maximizing displacement or for energy efficiency, as well as incorporating additional effects such as viscous drag, vortex shedding and hydrodynamic interaction into the theoretical models. It is also planned to experimentally investigate the dependence of motion on the actuation frequency, in order to test the model's assumption of time-invariant dynamics.



\section*{Acknowledgment}

We would like to warmly thank Prof. Tal Shima, head of the Technion's Laboratory of Cooperative Autonomous Systems for hosting the experiments at his lab. We thank Daniel Weinfeld, Sergey Shulman and Ilya Dakuko for arranging the experimental setup and for their help during the experiments. Special thanks to Ilya Dakuko for assembling the wireless operation system on the robots. We also thank Arik Bar-Yehuda and Noam Zriehan for constructing the pool, and Asaf Greenberg and Elon Tovi for designing and building the robot and for helping with the experiments. Finally, we wish to thank Atai Baldinger for his help with the experiments and their data analysis. 



\begin{thebibliography}{10}

\bibitem{bayat2016envirobot}
Behzad Bayat, Alessandro Crespi, and Auke Ijspeert.
\newblock Envirobot: A bio-inspired environmental monitoring platform.
\newblock In {\em Autonomous Underwater Vehicles (AUV), 2016 IEEE/OES}, pages
  381--386. IEEE, 2016.

\bibitem{cortes2001optimal}
Jorge Cort{\'e}s, Sonia Mart{\'\i}nez, James~P Ostrowski, and Kenneth~A
  McIsaac.
\newblock Optimal gaits for dynamic robotic locomotion.
\newblock {\em The International Journal of Robotics Research}, 20(9):707--728,
  2001.

\bibitem{crespi2008online}
Alessandro Crespi and Auke~Jan Ijspeert.
\newblock Online optimization of swimming and crawling in an amphibious snake
  robot.
\newblock {\em IEEE Transactions on Robotics}, 24(1):75--87, 2008.

\bibitem{crespi2013salamandra}
Alessandro Crespi, Konstantinos Karakasiliotis, Andre Guignard, and Auke~Jan
  Ijspeert.
\newblock Salamandra robotica ii: an amphibious robot to study salamander-like
  swimming and walking gaits.
\newblock {\em IEEE Transactions on Robotics}, 29(2):308--320, 2013.

\bibitem{gong2016kinematic}
Chaohui Gong, Matthew~J Travers, Henry~C Astley, Lu~Li, Joseph~R Mendelson,
  Daniel~I Goldman, and Howie Choset.
\newblock Kinematic gait synthesis for snake robots.
\newblock {\em The International Journal of Robotics Research},
  35(1-3):100--113, 2016.

\bibitem{gutman2016symmetries}
Emiliya Gutman and Yizhar Or.
\newblock Symmetries and gaits for {P}urcell's three-link microswimmer model.
\newblock {\em IEEE Transactions on Robotics}, 32(1):53--69, 2016.

\bibitem{hatton2011geometric}
Ross~L Hatton and Howie Choset.
\newblock Geometric motion planning: The local connection, {S}tokes' theorem,
  and the importance of coordinate choice.
\newblock {\em The International Journal of Robotics Research},
  30(8):988--1014, 2011.

\bibitem{hatton2013geometric}
Ross~L Hatton and Howie Choset.
\newblock Geometric swimming at low and high {R}eynolds numbers.
\newblock {\em IEEE Transactions on Robotics}, 29(3):615--624, 2013.

\bibitem{hatton2015nonconservativity}
Ross~L Hatton and Howie Choset.
\newblock Nonconservativity and noncommutativity in locomotion.
\newblock {\em The European Physical Journal Special Topics},
  224(17-18):3141--3174, 2015.

\bibitem{hirose1993biologically}
Shigeo Hirose.
\newblock Biologically inspired robot.
\newblock {\em Oxford University Press}, 1993.

\bibitem{ijspeert2014biorobotics}
Auke~J Ijspeert.
\newblock Biorobotics: Using robots to emulate and investigate agile
  locomotion.
\newblock {\em Science}, 346(6206):196--203, 2014.

\bibitem{kanso2005locomotion}
Eva Kanso, Jerrold~E Marsden, Clarence~W Rowley, and Juan~B Melli-Huber.
\newblock Locomotion of articulated bodies in a perfect fluid.
\newblock {\em Journal of Nonlinear Science}, 15(4):255--289, 2005.

\bibitem{kelasidi2015experimental}
Eleni Kelasidi, P{\aa}l Liljeb{\"a}ck, Kristin~Y Pettersen, and Jan~T Gravdahl.
\newblock Experimental investigation of efficient locomotion of underwater
  snake robots for lateral undulation and eel-like motion patterns.
\newblock {\em Robotics and biomimetics}, 2(1):8, 2015.

\bibitem{kelasidi2016innovation}
Eleni Kelasidi, Pal Liljeback, Kristin~Y Pettersen, and Jan~Tommy Gravdahl.
\newblock Innovation in underwater robots: biologically inspired swimming snake
  robots.
\newblock {\em IEEE Robotics \& Automation Magazine}, 23(1):44--62, 2016.

\bibitem{kelasidi2014modeling}
Eleni Kelasidi, Kristin~Ytterstad Pettersen, and Jan~Tommy Gravdahl.
\newblock Modeling of underwater snake robots moving in a vertical plane in 3d.
\newblock In {\em Intelligent Robots and Systems (IROS 2014), 2014 IEEE/RSJ
  International Conference on}, pages 266--273. IEEE, 2014.

\bibitem{kelly1995geometric}
Scott~D Kelly and Richard~M Murray.
\newblock Geometric phases and robotic locomotion.
\newblock {\em Journal of Robotic Systems}, 12(6):417--431, 1995.

\bibitem{kwak2016design}
Bokeon Kwak and Joonbum Bae.
\newblock Design of a robot with biologically-inspired swimming hairs for fast
  and efficient mobility in aquatic environment.
\newblock In {\em Intelligent Robots and Systems (IROS), 2016 IEEE/RSJ
  International Conference on}, pages 4970--4975. IEEE, 2016.

\bibitem{lamb1945hydrodynamics}
Horace Lamb.
\newblock {\em Hydrodynamics}, volume~43.
\newblock Dover, New York, 1945.

\bibitem{lee2009dynamics}
Taeyoung Lee, Melvin Leok, and N~Harris McClamroch.
\newblock Dynamics of connected rigid bodies in a perfect fluid.
\newblock In {\em American Control Conference, 2009. ACC'09.}, pages 408--413.
  IEEE, 2009.

\bibitem{li2016bio}
Guoyuan Li, Yuxiang Deng, Ottar~L Osen, Shusheng Bi, and Houxiang Zhang.
\newblock A bio-inspired swimming robot for marine aquaculture applications:
  From concept-design to simulation.
\newblock In {\em OCEANS 2016-Shanghai}, pages 1--7. IEEE, 2016.

\bibitem{mcisaac2002experimental}
Kenneth~A McIsaac and James~P Ostrowski.
\newblock Experimental verification of open-loop control for an underwater
  eel-like robot.
\newblock {\em The International Journal of Robotics Research},
  21(10-11):849--859, 2002.

\bibitem{melli2006motion}
Juan~B Melli, Clarence~W Rowley, and Dzhelil~S Rufat.
\newblock Motion planning for an articulated body in a perfect planar fluid.
\newblock {\em SIAM Journal on Applied Dynamical Systems}, 5(4):650--669, 2006.

\bibitem{morgansen2002trajectory}
Kristi~A Morgansen, Patricio~A Vela, and Joel~W Burdick.
\newblock Trajectory stabilization for a planar carangiform robot fish.
\newblock In {\em Robotics and Automation, 2002. Proceedings. ICRA'02. IEEE
  International Conference on}, volume~1, pages 756--762. IEEE, 2002.

\bibitem{nayfeh2008perturbation}
Ali~H Nayfeh.
\newblock {\em Perturbation methods}.
\newblock John Wiley \& Sons, 2008.

\bibitem{onal2013autonomous}
Cagdas~D Onal and Daniela Rus.
\newblock Autonomous undulatory serpentine locomotion utilizing body dynamics
  of a fluidic soft robot.
\newblock {\em Bioinspiration \& biomimetics}, 8(2):026003, 2013.

\bibitem{ostrowski2000optimal}
James~P Ostrowski, Jaydev~P Desai, and Vijay Kumar.
\newblock Optimal gait selection for nonholonomic locomotion systems.
\newblock {\em The International journal of robotics research}, 19(3):225--237,
  2000.

\bibitem{ostrowski1996gait}
Jim Ostrowski and Joel Burdick.
\newblock Gait kinematics for a serpentine robot.
\newblock In {\em Robotics and Automation, 1996. Proceedings., 1996 IEEE
  International Conference on}, volume~2, pages 1294--1299. IEEE, 1996.

\bibitem{ostrowski1998geometric}
Jim Ostrowski and Joel Burdick.
\newblock The geometric mechanics of undulatory robotic locomotion.
\newblock {\em The international Journal of Robotics Research}, 17(7):683--701,
  1998.

\bibitem{porez2014improved}
Mathieu Porez, Fr{\'e}d{\'e}ric Boyer, and Auke~Jan Ijspeert.
\newblock Improved {L}ighthill fish swimming model for bio-inspired robots:
  Modeling, computational aspects and experimental comparisons.
\newblock {\em The International Journal of Robotics Research},
  33(10):1322--1341, 2014.

\bibitem{ramasamy2017geometric}
Suresh Ramasamy and Ross~L Hatton.
\newblock Geometric gait optimization beyond two dimensions.
\newblock In {\em American Control Conference (ACC), 2017}, pages 642--648.
  IEEE, 2017.

\bibitem{sfakiotakis2007biomimetic}
Michael Sfakiotakis and Dimitris~P Tsakiris.
\newblock Biomimetic centering for undulatory robots.
\newblock {\em The International Journal of Robotics Research},
  26(11-12):1267--1282, 2007.

\bibitem{shammas2007geometric}
Elie~A Shammas, Howie Choset, and Alfred~A Rizzi.
\newblock Geometric motion planning analysis for two classes of underactuated
  mechanical systems.
\newblock {\em The International Journal of Robotics Research},
  26(10):1043--1073, 2007.

\bibitem{tallapragada2015self}
P~Tallapragada and SD~Kelly.
\newblock Self-propulsion of free solid bodies with internal rotors via
  localized singular vortex shedding in planar ideal fluids.
\newblock {\em The European Physical Journal Special Topics},
  224(17-18):3185--3197, 2015.

\bibitem{tam2007optimal}
Daniel Tam and Annete~E Hosoi.
\newblock Optimal stroke patterns for {P}urcell's three-link swimmer.
\newblock {\em Physical Review Letters}, 98(6):068105, 2007.

\bibitem{vaidyanathan2000hydrostatic}
Ravi Vaidyanathan, Hillel~J Chiel, and Roger~D Quinn.
\newblock A hydrostatic robot for marine applications.
\newblock {\em Robotics and Autonomous Systems}, 30(1):103--113, 2000.

\bibitem{wiezel2018energy}
Oren Wiezel, Laetitia Giraldi, Antonio DeSimone, Yizhar Or, and Fran{\c{c}}ois
  Alouges.
\newblock Energy-optimal small-amplitude strokes for multi-link microswimmers:
  {P}urcell's loops and {T}aylor's waves reconciled.
\newblock {\em arXiv preprint arXiv:1801.04687}, 2018.

\bibitem{wiezel2016optimization}
Oren Wiezel and Yizhar Or.
\newblock Optimization and small-amplitude analysis of {P}urcell's three-link
  microswimmer model.
\newblock {\em Proc. R. Soc. A}, 472(2192):20160425, 2016.

\bibitem{wiezel2016using}
Oren Wiezel and Yizhar Or.
\newblock Using optimal control to obtain maximum displacement gait for
  {P}urcell's three-link swimmer.
\newblock In {\em Decision and Control (CDC), 2016 IEEE 55th Conference on},
  pages 4463--4468. IEEE, 2016.

\bibitem{zhu2002three}
Q~Zhu, MJ~Wolfgang, DKP Yue, and MS~Triantafyllou.
\newblock Three-dimensional flow structures and vorticity control in fish-like
  swimming.
\newblock {\em Journal of Fluid Mechanics}, 468:1--28, 2002.

\end{thebibliography}

\end{document}